\documentclass{jfm}

\usepackage{graphicx}
\graphicspath{{Plots/}}
\usepackage[singlelinecheck=false,justification=justified]{caption}
\usepackage{newtxtext}
\usepackage{newtxmath}
\usepackage{natbib}
\usepackage{hyperref}
\hypersetup{
    colorlinks = true,
    urlcolor   = blue,
    citecolor  = black,
}

\newcommand{\RomanNumeralCaps}[1]
\linenumbers
\usepackage[export]{adjustbox}
\usepackage{subcaption}
\usepackage{ragged2e}
\usepackage{siunitx}
\usepackage{xcolor}
\usepackage[section]{placeins}
\usepackage{float}  
\usepackage{physics} 
\usepackage{tikz,tkz-euclide}
\usetikzlibrary{shapes}
\usetikzlibrary{positioning}
\usepackage[normalem]{ulem}
\usepackage{color}
\usetikzlibrary{calc,patterns,angles,quotes,arrows}
\usepackage{bm}  
\usepackage{cleveref}
\Crefname{figure}{Fig.}{Figs.}
\Crefname{equation}{Eqn.}{Eqns.}
\usepackage{gensymb}
\usepackage{enumitem}

\usepackage{bm}% bold math

\definecolor{black}{RGB}{0,0,0}
\newcommand{\blackcirc}{\raisebox{0.5pt}{\tikz{\node[draw,scale=0.4,black,circle,fill=black](){};}}}

\definecolor{red}{RGB}{187,0,0}
\newcommand{\redtri}{\raisebox{0pt}{\tikz{\node[draw,scale=0.3,red,regular polygon, regular polygon sides=3,fill=red,rotate=180](){};}}}
\newcommand{\redtrinf}{\raisebox{0pt}{\tikz{\node[draw,scale=0.3,red,regular polygon, regular polygon sides=3,rotate=180](){};}}}

\definecolor{blue}{RGB}{34,34,255}
\newcommand{\bluesq}{\raisebox{0pt}{\tikz{\node[draw,scale=0.4,blue,regular polygon, regular polygon sides=4,fill=blue](){};}}}
\newcommand{\bluesqnf}{\raisebox{0pt}{\tikz{\node[draw,scale=0.4,blue,regular polygon, regular polygon sides=4](){};}}}

\definecolor{redl}{RGB}{255,136,136}

\definecolor{green}{RGB}{0,128,0}
\newcommand{\greendia}{\raisebox{0pt}{\tikz{\node[draw,scale=0.4,green,diamond,fill=green](){};}}}

\definecolor{green}{RGB}{0,128,0}
\newcommand{\greentri}{\raisebox{0pt}{\tikz{\node[draw,scale=0.3,green,regular polygon, regular polygon sides=3,fill=green,rotate=0](){};}}}

\title{Controlling secondary flows in Taylor-Couette flow using spanwise superhydrophobic surfaces}

\author{Vignesh Jeganathan\aff{1},
  Tala Shannak\aff{1},
  Kamran Alba\aff{2}
  \corresp{\email{kalba@central.uh.edu,}},
 \and Rodolfo Ostilla-M\'onico\aff{1,3}
 \corresp{\email{rodolfo.ostilla@uca.es}}}

\affiliation{\aff{1}Department of Mechanical Engineering, University of Houston, Houston 77004, USA
\aff{2}Department of Engineering Technology, University of Houston, Houston 77004, USA
\aff{3}Escuela de Ingenier\'ia, Universidad de C\'adiz, Spain}

\begin{document}
\maketitle

\begin{abstract}
Turbulent shear flows are abundant in geophysical and astrophysical systems and in engineering-technology applications. They are often riddled with large-scale secondary flows that drastically modify the characteristics of the primary stream, preventing or enhancing mixing, mass, and heat transfer. Using experiments and numerical simulations, we study the possibility of modifying these secondary flows by using superhydrophobic surface treatments that reduce the local shear. We focus on the canonical problem of Taylor-Couette flow, the flow between two coaxial and independently-rotating cylinders, which has robust secondary structures called Taylor rolls that persist even at significant levels of turbulence. We generate these structures by rotating only the inner cylinder of the system, and show that a spanwise superhydrophobic treatment can weaken the rolls through a mismatching surface heterogeneity, as long as the roll size can be fixed. The minimum hydrophobicity of the treatment required for this flow control is rationalized, and its effectiveness beyond the Reynolds numbers studied here is also discussed.
\end{abstract}


\section{Introduction}
\label{sec:intro}

 Shear flows are an extremely common occurrence in nature and technology. A simple example would be the fluid motion between two differentially moving parallel plates. More complex examples abound: from wind currents in the atmosphere at different speeds \citep{4362727,pedlosky1987geophysical}, to flow inside an industrial centrifugal reactor \citep{schrimpf2021taylor}. A primary shear flow field generally involves adjacent fluid layers that move at different speeds. Under certain conditions, the primary flow can be hydrodynamically unstable and any perturbation will lead to a complex three-dimensional flow, where secondary structures can arise that are superimposed on the primary stream \citep{orszag1983secondary}. Notable laminar secondary structures are found in the arteries in its curves and branches \citep{ku1997blood}. On the other hand, turbulent secondary structures are commonly found in geophysical and astrophysical occurrences, such as atmospheric convection cells responsible for the water cycle \citep{atkinson1996mesoscale,agee1984observations}. They also exist in centrifugal reactors \citep{schrimpf2021taylor} and rotating reverse osmosis filtration devices \citep{lee2001rotating,lee2001reverse}.

Secondary flows are a crucial component of the global properties of a system because they account for a major portion of the mass and momentum transport. Hence, the ability to affect or control these secondary structures could lead to affecting global transport properties or even the frictional losses in a system. The scientific interest behind this possibility has led to many attempts at secondary flow control \citep{bakhuis2018mixed,bakhuis2019controlling,naim2019turbulent,qi2012control}. A natural place to start this investigation is to simplify the flow as much as possible to canonical models. One of the most frequently studied canonical models for shear flows and its secondary structure is Taylor-Couette flow and its Taylor rolls, respectively.

Taylor-Couette flow (TCF) \citep{donnelly1991taylor,gro16} is the movement of the fluid between two concentric cylinders that rotate independently. A secondary flow called a Taylor roll forms if the flow is centrifugally (or Rayleigh) unstable, i.e.~if the angular momentum of the inner cylinder is larger than that of the outer cylinder. At low Reynolds numbers, these are axisymmetric and laminar \citep{tay23}. As the Reynolds number increases, they go through a series of instabilities where they transition to increasingly turbulent states: first to wavy Taylor vortex flow, then to  temporally modulated turbulent Taylor rolls, and finally to turbulent Taylor rolls \citep{andereck1986flow}. As large Reynolds numbers are reached ($Re\sim\mathcal{O}(10^5)$), turbulent Taylor rolls wash away in certain regions of parameter space, and where they remain, their main driver is the combination of shear and solid body rotation \citep{lathrop1992turbulent,huisman2014multiple,sacco2019dynamics}. 

Taylor rolls are a particularly interesting example of a secondary flow because of their robustness in the turbulent regime \citep{ostilla2017life,zhu2016direct}, and the possibility that multiple stable solutions (`roll states' with different roll sizes) can occur for the same boundary conditions \citep{coles1965transition,huisman2014multiple,martinez2014effect,wen2020controlling}. Furthermore, Taylor rolls are commonly used in engineering applications to affect mixing properties \citep{schrimpf2021taylor,lee2001rotating,lee2001reverse} and represent a large portion of the momentum transport across the cylinder gap \citep{brauckmann2013direct,ostilla2016near}.
By successfully modifying robust Taylor rolls, we can demonstrate a general capacity to modify secondary flows. The findings here can also be applied directly to TCF or TCF-like systems in engineering, such as centrifugal mixers or bioreactors. 

A successful approach to affect turbulent secondary structures is to force an additional secondary flow at a different wavelength from the existing structures. This would generate a mismatch that interferes destructively with existing or ``natural'' secondary flows \citep{bakhuis2018mixed,jeganathan2021controlling}. In the current study, we will attempt to induce Prandtl secondary flows of the second kind \citep{nikitin2021prandtl}, which are turbulent secondary flows that arise to compensate for imbalances (mainly in the mean Reynolds stresses) through turbulent pulsations and can be found, for example, in turbulent rectangular pipes \citep{hoagland1962fully}. The advantage of using this method is that destructive interference can be achieved through selective surface treating without substantially modifying an existing geometry. For example, patterns of heterogeneous roughness induce swirling motions in regions between high and low-momentum pathways \citep{nugroho2013large,barros2014observations,willingham2014turbulent,anderson2015numerical}. This swirling motion leads to secondary flows that are generated and sustained due to spanwise gradients in the Reynolds stress components, which cause an imbalance between the production and dissipation of turbulent kinetic energy that necessitates secondary advective velocities to balance \citep{anderson2015numerical}. In a similar spirit, a recent study has showed that heterogeneous spanwise roughness is a plausible mechanism to control secondary flows in TCF \citep{bakhuis2019controlling}. Through a combined use of experiments and simulations, this study has reported that certain distributions of roughness induce a new, spatially fixed secondary flow that is absent from the base flow, and this effect has resulted in a large change in flow properties. 

However, using roughness modifications to affect a system generally results in an increase in overall drag and causes energy losses in real-world applications. An alternative to using roughness, which increases local shear stress, is to attempt to induce the same types of secondary flow by using hydrophobic surfaces, which reduces local stress compared to untreated surfaces. This would induce similar stress imbalances and generate Prandtl secondary flows \citep{turk2014turbulent}. Idealized stress-free boundary inhomogeneities in TCF have been simulated in previous studies \citep{naim2019turbulent,jeganathan2021controlling}, which have reported a substantial modification of secondary flows when using spanwise (axial) boundary heterogeneity, with the effects persisting up to Reynolds numbers of the order of $Re\sim\mathcal{O}(10^4)$. The effects are greatest when the spanwise heterogeneity are distributed in a pattern with a characteristic wavelength of half the wavelength of the natural structure (a single Taylor roll), causing destructive interference between the two secondary flows \citep{jeganathan2021controlling}. 

A drawback of these numerical studies is that idealized stress-free boundaries are impossible to achieve in engineering applications, thus the potential for real-world applications is uncertain. In this manuscript, we set out to investigate whether this is experimentally feasible, i.e.~whether it is possible to control secondary flows using the types of stress-reducing surfaces available in a laboratory setting. To do this, we will use a highly-accessible superhydrophobic (SHP) coating \citep{jeevahan2018superhydrophobic} and assess its effects on TCF. Superhydrophobic coatings, unlike the stress-free limits attainable in computer simulations, have a finite slip length and are often heavily tested for their durability \citep{wang2016transparent,xue2015fabrication}. Recent efforts \citep{lambley2020superhydrophobic,wang2020design} show that it is indeed possible to achieve durable superhydrophobic surfaces that can withstand extreme conditions, paving the way for studies such as the present one.

The use of SHP surfaces for flow control has not been well explored, especially in the turbulent regime. However, there are some indications that they could be effective, such as reports that SHP surfaces can delay the onset of vortex shedding in flow over a cylinder, and also increase the shedding frequency of the Karman vortex, causing premature vortex roll up \citep{muralidhar2011influence,kim2015experimental}. Another study has pointed out the existence of a large number of coherent structures and a change in the vortex shedding pattern in the near wake of an SHP cylinder \citep{sooraj2020effect}. In addition to affecting secondary structures, the major impact of introducing stress-free or finite-slip boundary conditions is drag reduction. Indeed, our recent TCF simulations in  \cite{jeganathan2021controlling} reported torque reductions of up to 32\% when using (ideal) spanwise patterns of 50/50 no-slip/stress-free heterogeneity. In the laboratory, drag reduction through the use of SHP surfaces has been achieved in channel flow experiments \citep{wat99,ou2004laminar,tretheway2002apparent,cheng2002fluid}. TCF studies also report a maximum drag reduction of up to 80\% using chemically generated SHP surfaces \citep{srinivasan2015sustainable,rajappan2020cooperative,hu2017significant}, and up to 90\% using stress reduction limits generated by the Leidenfrost effect \citep{ayan2019experiments,saranadhi2016sustained}. However, we emphasize that a pure focus on drag reduction is not our main interest because through the application of SHP treatments, we expect to see a reduction in drag in the majority of cases, provided the surface is sufficiently hydrophobic and durable. We focus mainly on how the secondary structures are affected by these surface treatments, while making sure that the possible energy losses do not substantially increase.

\begin{figure}
\centering
\includegraphics[width=\linewidth]{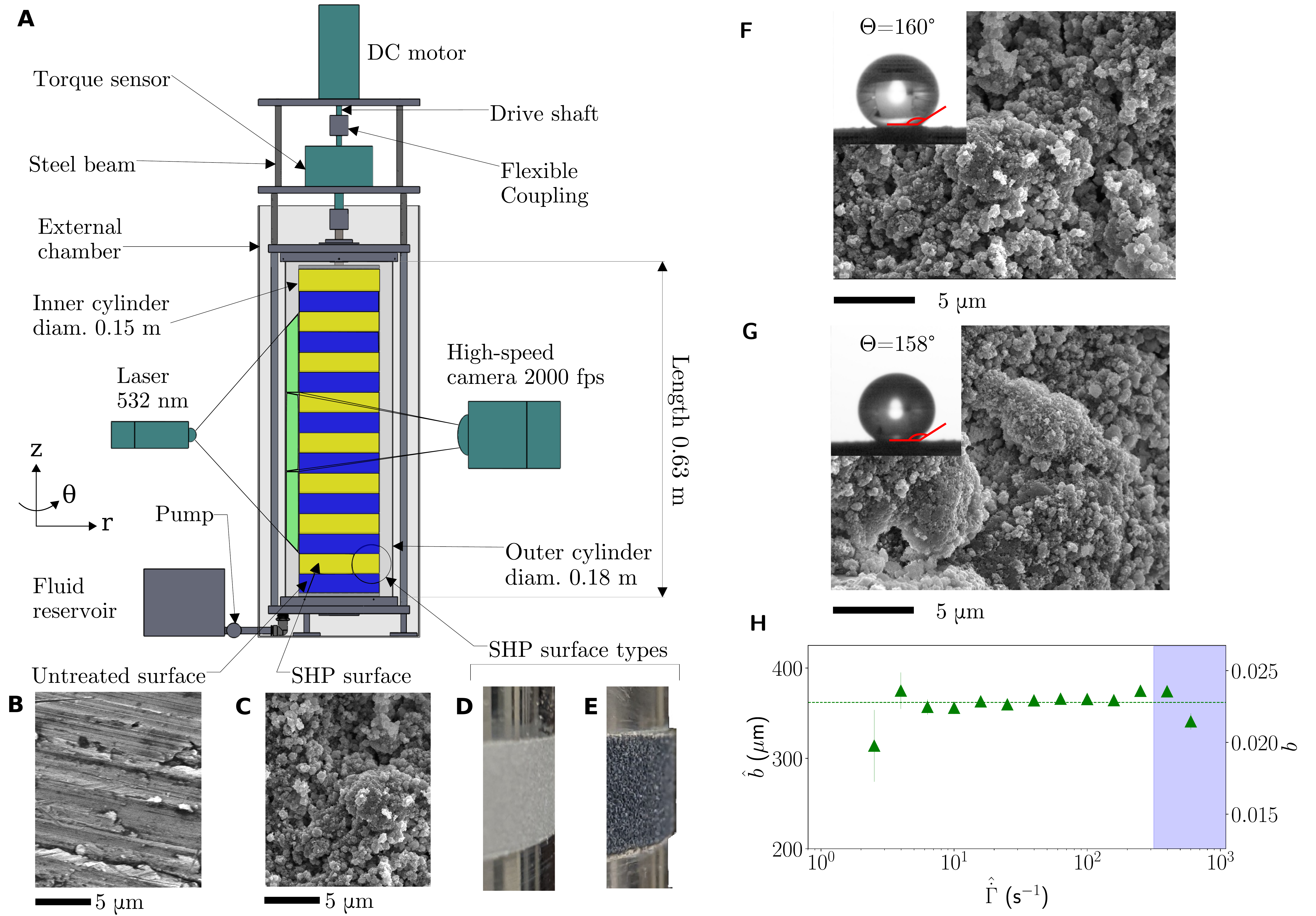}
\caption{(\textbf{A}) Schematic of experimental set-up. (\textbf{B, C}) Scanning Electron Microscope (SEM) images of the untreated aluminum surface (blue region in the schematic) and the superhydrophobic (SHP) surface (yellow region in the schematic), respectively. (\textbf{D}) Flat and (\textbf{E}) stepped SHP modifications on the inner cylinder. SEM image of (\textbf{F}) the freshly-coated SHP surface and (\textbf{G}) SHP surface sheared at the highest studied shear rate of $\hat{\dot{\Gamma}}=600$ s$^{-1}$ for 90 minutes. The insets in (\textbf{F}) and (\textbf{G}) depict the contact angle of a $5 \mu$L deionized water droplet on the corresponding surfaces. (\textbf{H}) Dimensional slip length \protect\greentri\ of the SHP coating measured using the rheometer at various tip shear rates. The right-hand $y$ axis shows the slip length non-dimensionalized by the cylinder gap width, $\hat d$, from the experiments. The dotted green line shows the average slip length $b=0.023$ of all tip shear rates. The error bars show the average of slip lengths measured during three separate instances \citep{srinivasan2013drag}. The blue region shows the zone of uncertainty caused by the onset of turbulence in rheometry tests; see Appendix \ref{sec:rheo_sup}}
  \label{fig:sch}
\end{figure}

To keep the parameter space simple, we apply SHP surface treatments only to the inner cylinder and focus on the resulting flow organization and torque for a TCF system with pure inner cylinder rotation. We study the Reynolds number in the order of $Re\sim\mathcal{O}(10^4)$, where the flow is turbulent and the Taylor rolls persist. Parameter space is further restricted to only axial (spanwise) pattern wavelengths that are large enough to have an impact on large-scale structures, rather than small pattern wavelengths that do not have a significant effect on the flow \citep{jeganathan2021controlling} and are more difficult to construct.

\section{Experiments}\label{sec:exp_sec}

\subsection{Experimental Methods}
A schematic of the experimental set-up is shown in \Cref{fig:sch}A. The dimensional and dimensionless parameters are consistently denoted with and without hat symbol, $\hat{}$, respectively throughout this manuscript. The Taylor-Couette experimental set-up is built using an aluminum inner cylinder of radius, $\hat r_i=76.2$ mm, and an acrylic outer cylinder of radius, $\hat r_o=92.1$ mm, leaving a gap width of $\hat d=\hat r_o-\hat r_i=15.9$ mm. The length of the cylinders is $\hat l=614.7$ mm. The resulting dimensionless geometric parameters are the radius ratio, $\eta=\hat r_i/\hat r_o=0.83$, and the aspect ratio, $\Gamma_z=\hat{l}/\hat{d}=38.7$. The outer cylinder is fixed, and the inner cylinder is rotated at a rotational velocity, $\hat \omega_i$, driven by a brushless DC motor (IKA Eurostar 200 mixer). The shear driving strength can be expressed as a Reynolds number of the inner cylinder, $Re_i=\hat r_i \hat \omega_i \hat d/\hat{\nu}$, where $\hat \nu$ is the kinematic viscosity of the working fluid. 

As shown in \Cref{fig:sch}A, there is a small gap between the inner cylinder and the end caps at the top and bottom. The end caps are stationary, and this would mean a discontinuity in velocity between the inner cylinder and end caps. To minimize torque losses, the system is filled in a way such that the fluid only fills the gap up to the top surface of the inner cylinder. This means that only the space between the inner cylinder and the bottom end cap contains liquid. The gap between the inner cylinder and the cylinder's top cap contains air. With this set-up we estimate that 20-30\% of the measured torque results from the end caps and other system losses. 
 
% \begin{figure}
% \centering
% \includegraphics[width=0.8\linewidth]{fig4_2}
% \caption{ Superhydrophobic (SHP) surface characteristics. SEM image of (\textbf{A}) the freshly-coated SHP surface and (\textbf{B}) SHP surface sheared at the highest studied shear rate of $\hat{\dot{\Gamma}}=600$ s$^{-1}$ for 90 minutes. The insets in (\textbf{A}) and (\textbf{B}) depict the contact angle of a $5 \mu$L deionized water droplet on the corresponding surfaces. (\textbf{C}) Dimensional slip length \protect\greentri\ of the SHP coating measured using the rheometer at various shear rates. The right-hand $y$ axis shows the slip length non-dimensionalized by the cylinder gap width, $\hat d$, from the experiments. The dotted green line shows the average slip length $b=0.023$ of all shear rates. The error bars show the average of slip lengths measured during three separate instances \citep{srinivasan2013drag}. The blue region shows the zone of uncertainty caused by the onset of turbulence in rheometry tests; see Supplementary Information for more details.}
%   \label{fig:surf}
% \end{figure}

To achieve the SHP surfaces required to construct TCF experiments, we use a commercial two-step coating called Ultra-Ever Dry, UltraTech International, as used in a previous TCF study by \cite{hu2017significant}. The first step requires spraying a chemical called `bottom coat' followed by the `top coat' in the second step. The bottom coat is not superhydrophobic, but once it cures, it facilitates the self-assembly and bonding characteristic of the microstructures responsible for the superhydrophobicity found in the `top coat'. Microscope images comparing the uncoated and SHP surfaces are shown in Figs.~\ref{fig:sch} B and C, respectively. We can clearly see that the SHP-treated surface has air-trapping microstructures that cause superhydrophobicity which are largely absent on the uncoated surface. There are two ways by which one may apply the SHP coating on the inner cylinder in TCF experiments. In the first method, the treated surfaces are made by sandblasting the inner cylinder, followed by spraying the coatings. This method leaves a nearly-flat surface on the inner cylinder as shown in \Cref{fig:sch}D. Therefore, we refer to this treatment as `flat SHP'. In the second method, we spray the SHP coating on abrasive tapes of 80 grit size, achieving a combined thickness of tape and coating of $0.58$ mm. These are fixed on the inner cylinder using an adhesive backing. This leaves a step-like structure on the system as shown in \Cref{fig:sch}E. We accordingly refer to this treatment as `stepped SHP'. SHP patterns are applied to the inner cylinder in an axially periodic manner, as shown in \Cref{fig:sch}A. This is achieved by masking portions of the inner cylinder during the coating process.

We define the dimensionless SHP pattern wavelength as $\lambda_z= 2\hat s_z/\hat d$, with $\hat s_z$ being the dimensional axial width of the coating. We will explore three pattern sizes:  $\lambda_z=1.2$, $\lambda_z=2.4$ and  $\lambda_z=4.8$. We have chosen these three values because of several reasons: (i) they serve to divide the cylinder equally, (ii) they roughly correspond to the values of $\lambda_z$ studied in \cite{jeganathan2021controlling} ($1.17$, $2.33$ and $4.6$) and (iii) they roughly correspond to one-half, one or twice the Taylor roll wavelength one can expect at these Reynolds numbers. This last point is important, as it can serve to experimentally test the prediction from \cite{jeganathan2021controlling}, where we found that $\lambda_z=\frac{1}{2}\lambda_{TR}$ was the most effective wavelength in disrupting the turbulent Taylor rolls. We note that with this choice, we cannot distinguish whether $\lambda_z=1.2$, $\lambda_z=1.3$ or $\lambda=1.4$ would be the best fit for this (or any) Taylor-Couette system, but instead focus on giving a proof of concept that axial heterogeneities with wavelengths similar to half the roll size can disrupt turbulent Taylor rolls in an experiment, and that they work better than axial heterogeneities at wavelengths comparable to a single or double roll size. Finally, as mentioned in the introduction, we did not study patterns with wavelengths smaller than $\approx \frac{1}{2}\lambda_{TR}$, as we do not expect them to affect the rolls substantially \citep{jeganathan2021controlling}.

To depict the SHP microstructures more clearly, we show the microscopic image of a freshly-coated SHP surface in \Cref{fig:sch}F. These microstructures display superhydrophobicity by creating a low surface energy \citep{jeevahan2018superhydrophobic} and causing the droplet contact angle to be as high as $\Theta=160\degree \pm 2\degree$ as shown in the insert of \Cref{fig:sch}F. To demonstrate the durability of this coating, we present a microscopic image of a water droplet and its contact angle with an SHP surface which has been sheared for 90 minutes at a shear rate of $\hat{\dot{\Gamma}}=600$ s$^{-1}$, in \Cref{fig:sch}G. We use this shear rate and duration as they correspond to the highest $Re_i$ in this study, $Re_i=2\times10^4$ and to the time frame of the TCF experiments. The sheared surface still retains its SHP microstructures and a high contact angle of $\Theta=158\degree$, giving us confidence in the ability of the coating to withstand TCF experiments. To study the stress-reducing characteristics of SHP surface, we use the method detailed in Ref.~\citep{srinivasan2013drag}. The experimental slip length at different tip shear rates $2$ s$^{-1}<\hat{\dot{\Gamma}}<600$ s$^{-1}$ is derived from rheometer (HR-3 Discovery hybrid model, TA Instruments) measurements presented in \Cref{fig:sch}H. The average slip length across all the shear rates is found out to be $\hat b=360 \pm 12\ \mu $m. This corresponds to $b=\hat{b}/\hat{d}=0.023 $ in dimensionless terms. Further details of the characterization of the SHP surface are provided in the Appendices \ref{sec:rheo_sup} and \ref{sec:sh_test}. 

Once the SHP surfaces are fixed to the inner cylinders of different axial patterns, we start a series of TCF experiments. The gap between the inner and outer cylinders is filled with deionized water, which is seeded with $50\ \mu$m polyamide seeding particle at 0.2 g/L to obtain particle image velocity (PIV) data \citep{buchhave1992particle}. The working fluid is isolated between the cylinders using various rotary and static rubber seals. Temperature fluctuations are recorded using an Omega HH308 thermometer, revealing that they are within $0.1$ K during the PIV experiments. LaVision Nd:YAG laser ($532$ nm) is used to generate a vertical laser sheet of thickness $2$ mm that illuminates the gap between the cylinders. To reduce light refraction errors from the curved acrylic outer cylinder, we have placed the TCF cell inside a cuboidal external chamber. The cuboidal chamber is made from acrylic and filled with water that has a refractive index close to that of acrylic, creating a \emph{fish tank} \citep{moises2016isodense}. 

The system is started up by accelerating the inner cylinder at 0.279 rad/s$^2$ to reach the desired rpm. For most cases, this achieves a reliable number of rolls, as we detail below. Before starting the PIV measurements, we wait five minutes at a given $Re_i$ so that the flow achieves a statistically stationary state. After this period, we use a high-speed camera (Phantom VEO 710) to record 6000 PIV images (1280$\times$504 pixels) of the fully-developed Taylor rolls at $2000$ fps for a time period of 3 s. We selected this fps after an extensive parametric study to achieve high-resolution images. The 3-s time period corresponds to $\approx 300$ eddy turnover times of Taylor rolls, $\hat{t}_e =  \hat{d}/(\hat{r}_i\hat{\omega}_i)$, which is long enough to study their properties, while the initial wait of five minutes corresponds to $\approx 3\cdot10^5$ large-eddy turnover times, more than enough to achieve a statistically stationary state \citep{ostilla2016near}. The camera is fitted with a K2 DistaMax long-distance microscope to achieve a 4.4x zoom factor. The PIV images are post-processed in MATLAB's open-source PIVlab software \citep{thielicke2021particle} using multi-step interrogation windows ranging from $64 \times 64$ to $32 \times 32$ pixels. We then obtain the instantaneous radial, $\hat{u}_r$, and axial, $\hat{u}_z$, velocity components of the flow. Torque, $\hat T$, is measured for $600$ s at a rate of $1000$ Hz using an inline rotary ultra-precision torque sensor (Himmelstein MCRT 48801V[25-0]CFZ). The torque sensor is attached to the driving shaft that connects the motor to the TCF cell. In Appendix \ref{sec:expcomp}, we compare velocity and torque data obtained from our experiment to other experiments and simulations. We use low internal clearance P5 high-precision deep groove SKF ball bearings to reduce the effects of frictional force on the rotary seals, centrifugal forces, and the buoyancy of the rotating inner cylinder on the measured torque. The temperature of water is measured during the torque measurement experiments, and the corresponding viscosities and densities are used in the Reynolds number calculations. The density and viscosity of the working fluids at different temperatures are measured using a hand-held density meter (DMA Basic 35)  and a Rheometer (HR-3 Discovery hybrid model, TA Instruments) respectively.

\subsection{Experimental Results}
\label{sec:expresults}

To visualize the classic no-slip turbulent TCF, we present the temporally-averaged dimensionless radial velocity $\langle u_r \rangle_t$ of the flow field at $Re_i=10^4$ in \Cref{fig:exp_vs}A and the corresponding PIV experiment in Movie M1 of Appendix \ref{sec:vid}. The $x$ and $y$ axes are nondimensionalized by gap width, $\hat d$, with $r=0$ corresponding to the inner cylinder and $r=1$ the outer cylinder. The average radial velocity $\langle \hat u_r \rangle_t$ is non-dimensionalized using the rotational velocity $\hat{r_i} \hat \omega_i$ of the inner cylinder to obtain the dimensionless velocity presented in \Cref{fig:exp_vs}. The effect of different wavelengths of the SHP pattern on the turbulent TCF flow field for both flat and stepped patterns is shown in \Cref{fig:exp_vs} at $Re_i=10^4$. For the flat patterns, it is apparent that the size of the rolls formed in the system changes across different patterns, something we can attribute to the Taylor-Couette system having a range of possible solutions reflected as changing roll wavelengths \citep{huisman2014multiple,martinez2014effect}. This finding is consistent with \cite{bakhuis2019controlling}, who has used axially-varying roughness to control a Taylor-Couette flow and has observed different roll sizes in their system. 

\begin{figure*}
\centering
\includegraphics[width=0.45\linewidth]{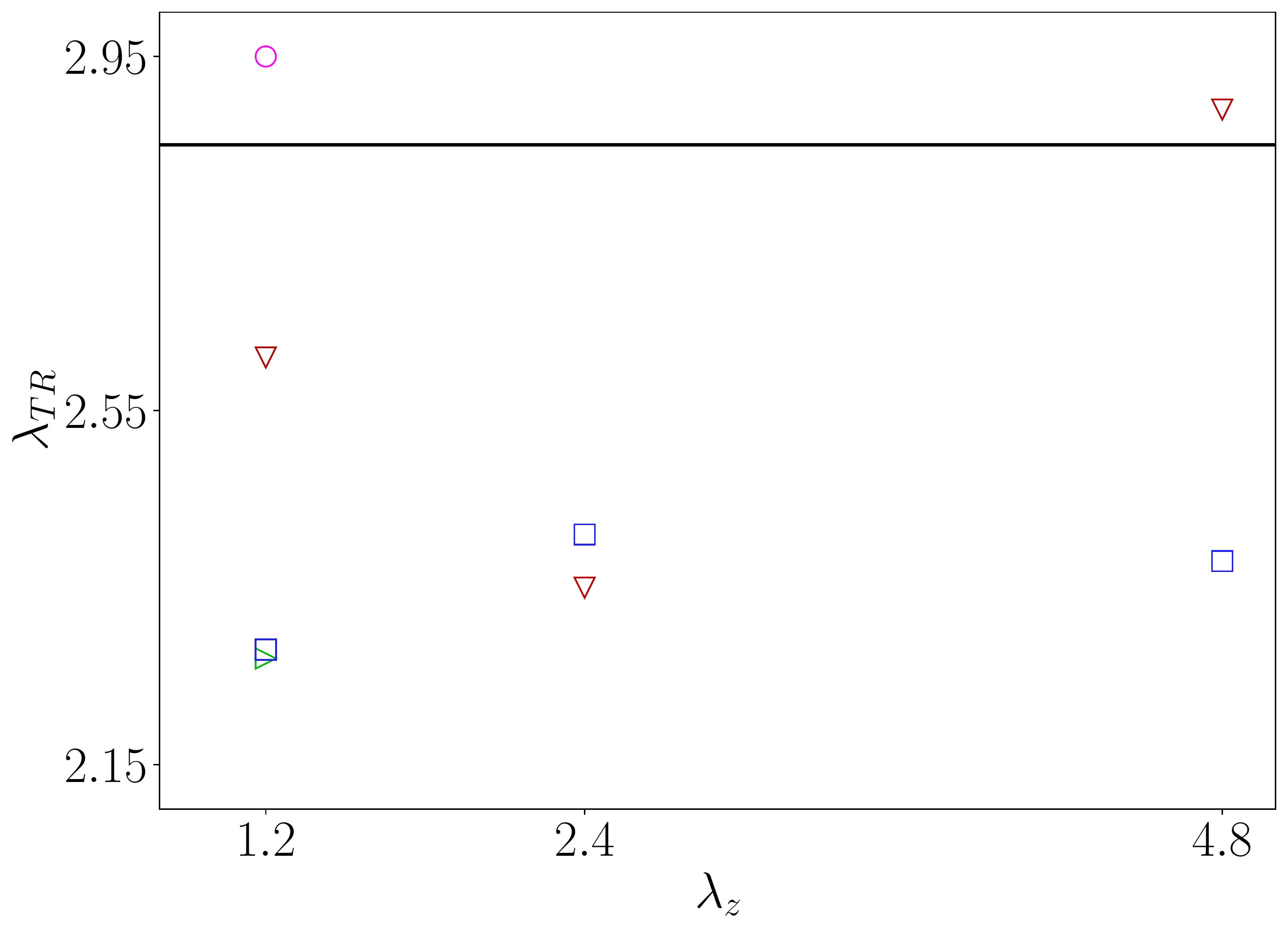}
\includegraphics[width=0.45\linewidth]{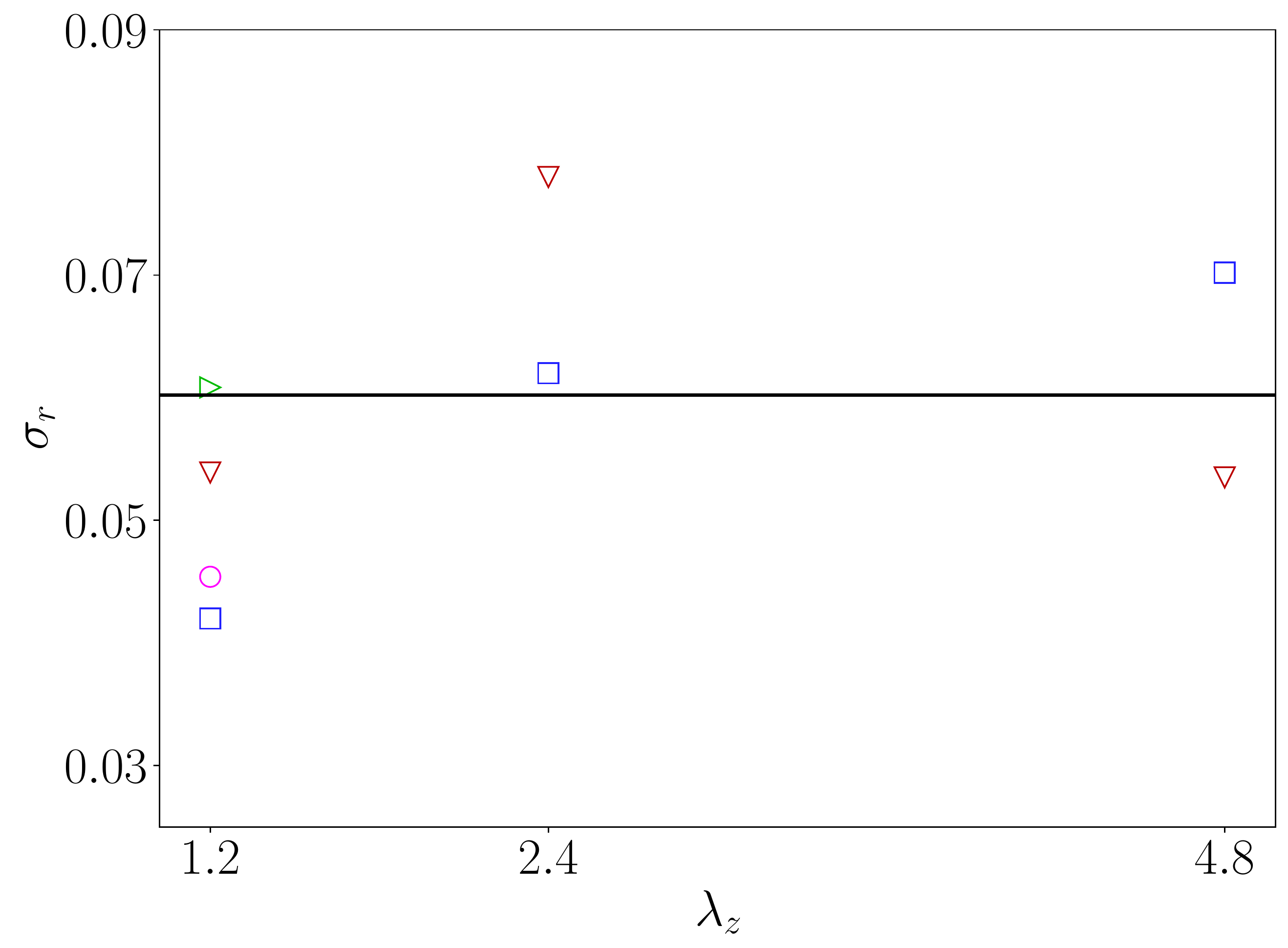}
\caption{Left: roll wavelength $\lambda_{TR}$ obtained from autocorrelations, for various flat \protect\redtrinf\ and stepped \protect\bluesqnf\ SHP pattern wavelengths at $Re_i=10^4$. Right: Roll strength $\sigma_r$ for the same cases. The black horizontal lines in represent the reference no-slip roll wavelength $\lambda_{TR}=2.85$ and roll strength $\sigma_r=6.02\times 10^{-2}$, respectively. The \textcolor{green}{$\rhd$} symbol represents a different roll state obtained for a flat pattern during experiments, and the \textcolor{magenta}{$\circ$} represents the stepped experiment with no treatment.}
  \label{fig:exp_trsigma}
\end{figure*}

To set the boundary of a pair of rolls and quantify the roll strength, we first calculate the roll (pair) wavelength $\lambda_{TR}$. This is given by twice the roll size determined from the axial autocorrelation of the radial velocity at the mid-gap, and in turn this is given by the (axial) distance to the first local minimum of the autocorrelation \citep{ostilla2015effects}. The left panel of \Cref{fig:exp_trsigma} ( \protect\redtrinf\ line ) shows how the different treatments affect $\lambda_{TR}$, quantifying the previous visual intuition: the roll wavelength $\lambda_{TR}$ varies by $\pm$20-30\% with respect to the no-slip reference case as the underlying pattern wavelength $\lambda_z$ is changed.

\begin{figure*}
\centering
 \includegraphics[height=0.3\linewidth]{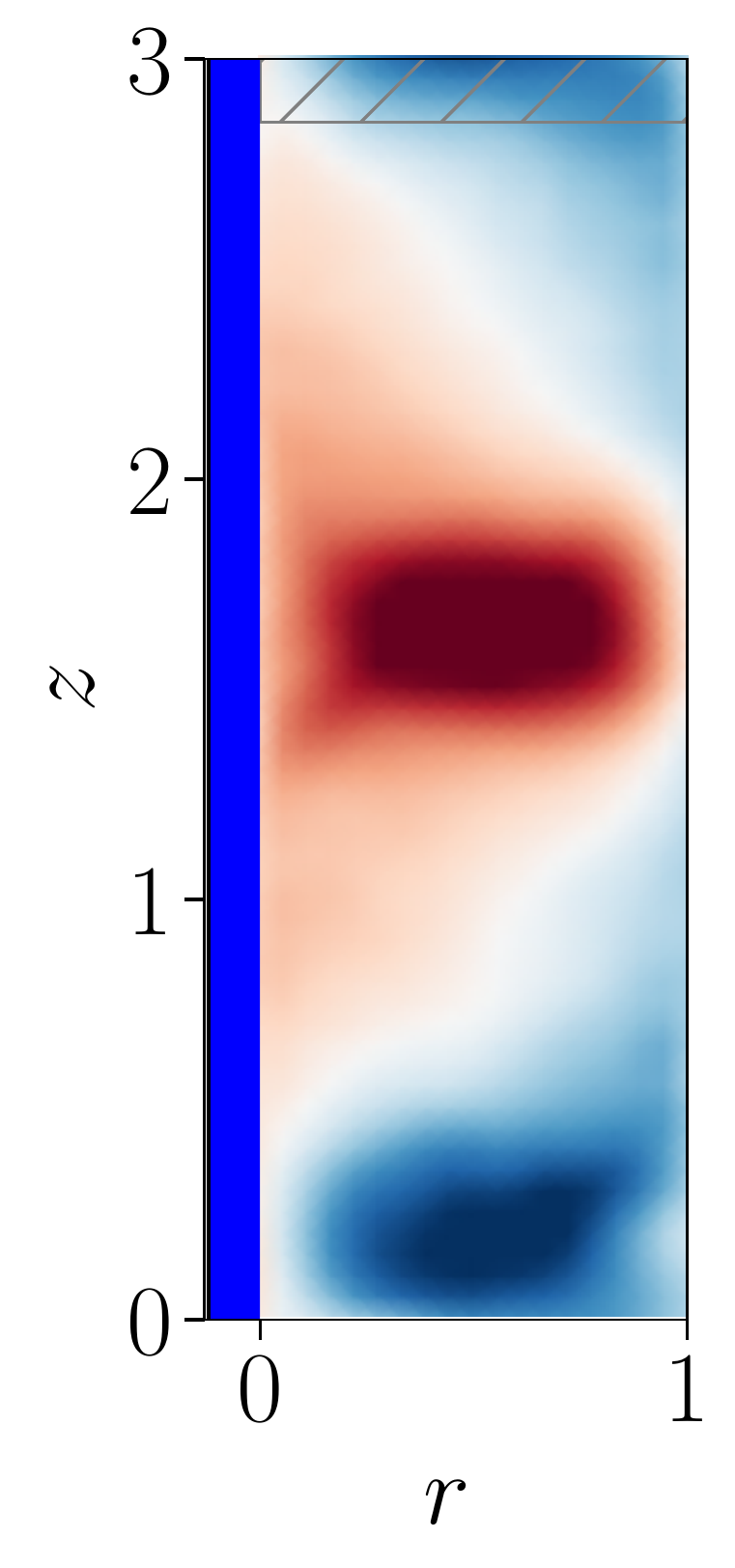}
  \includegraphics[height=0.3\linewidth]{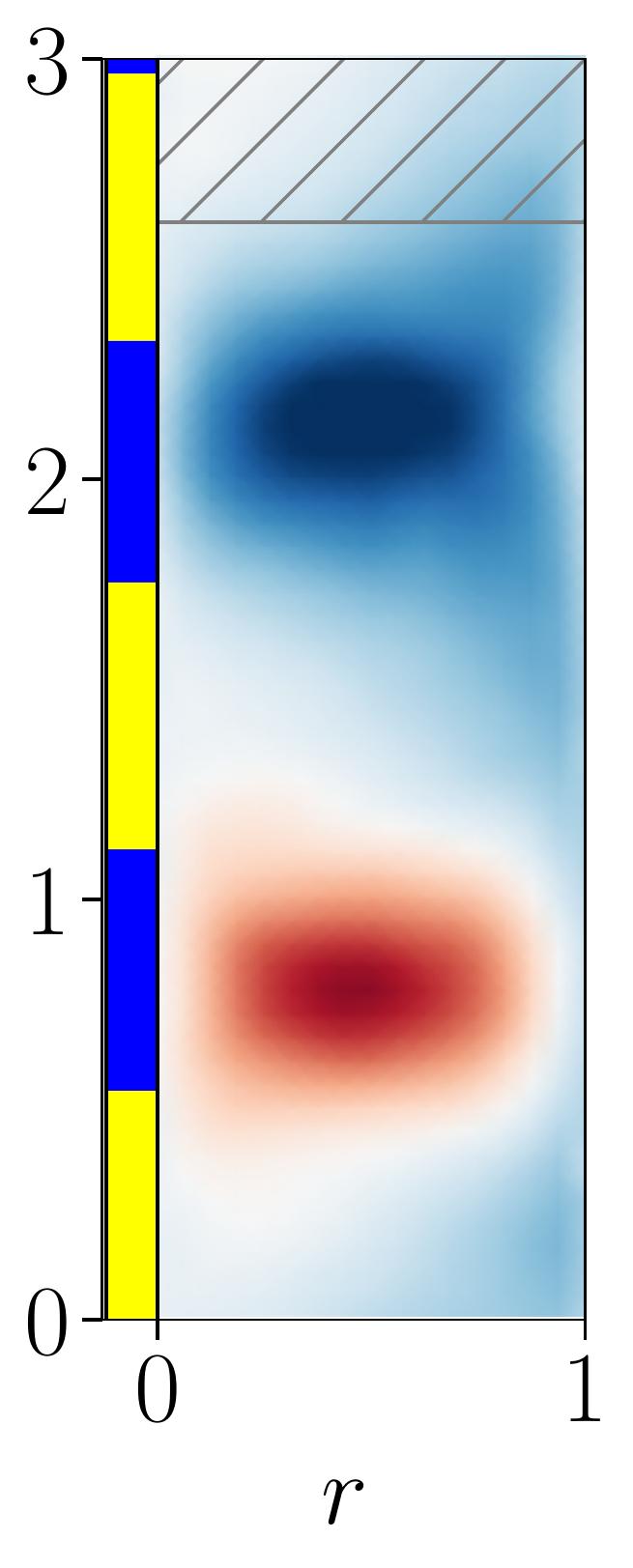}
 \includegraphics[height=0.3\linewidth]{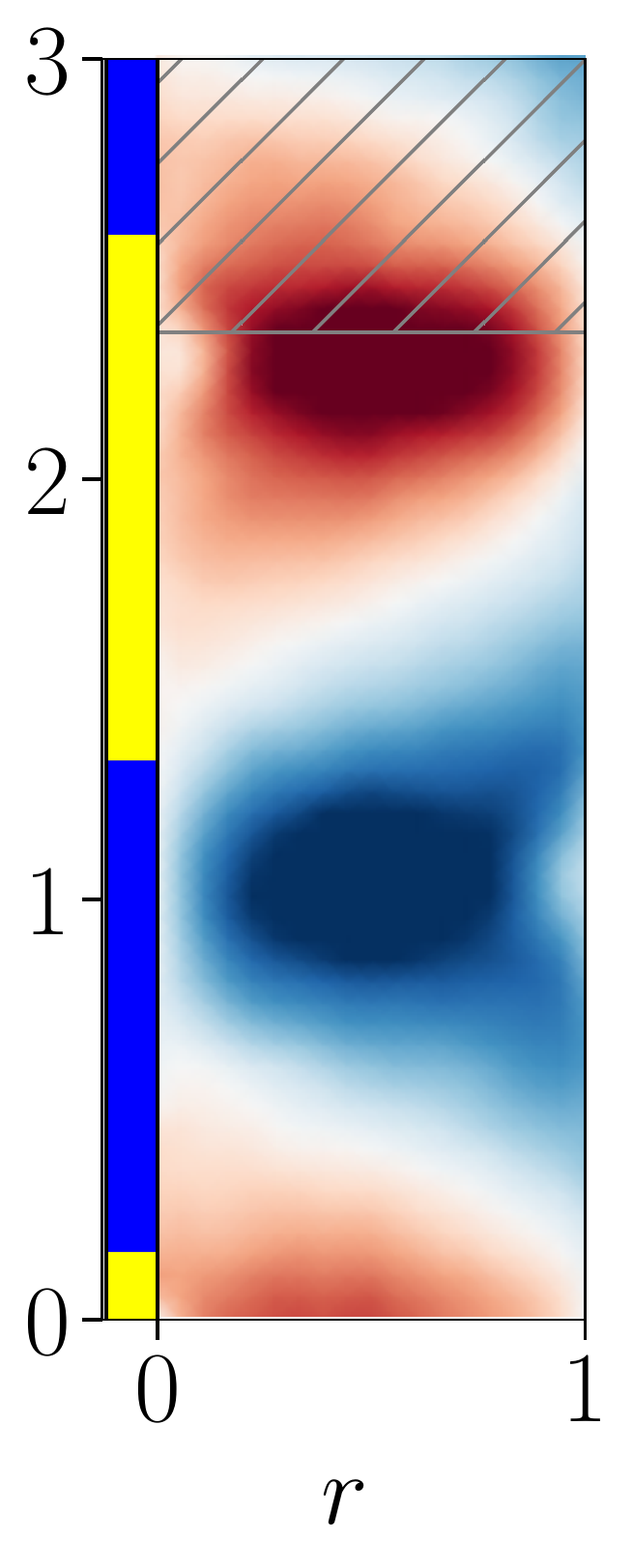}
 \includegraphics[height=0.3\linewidth]{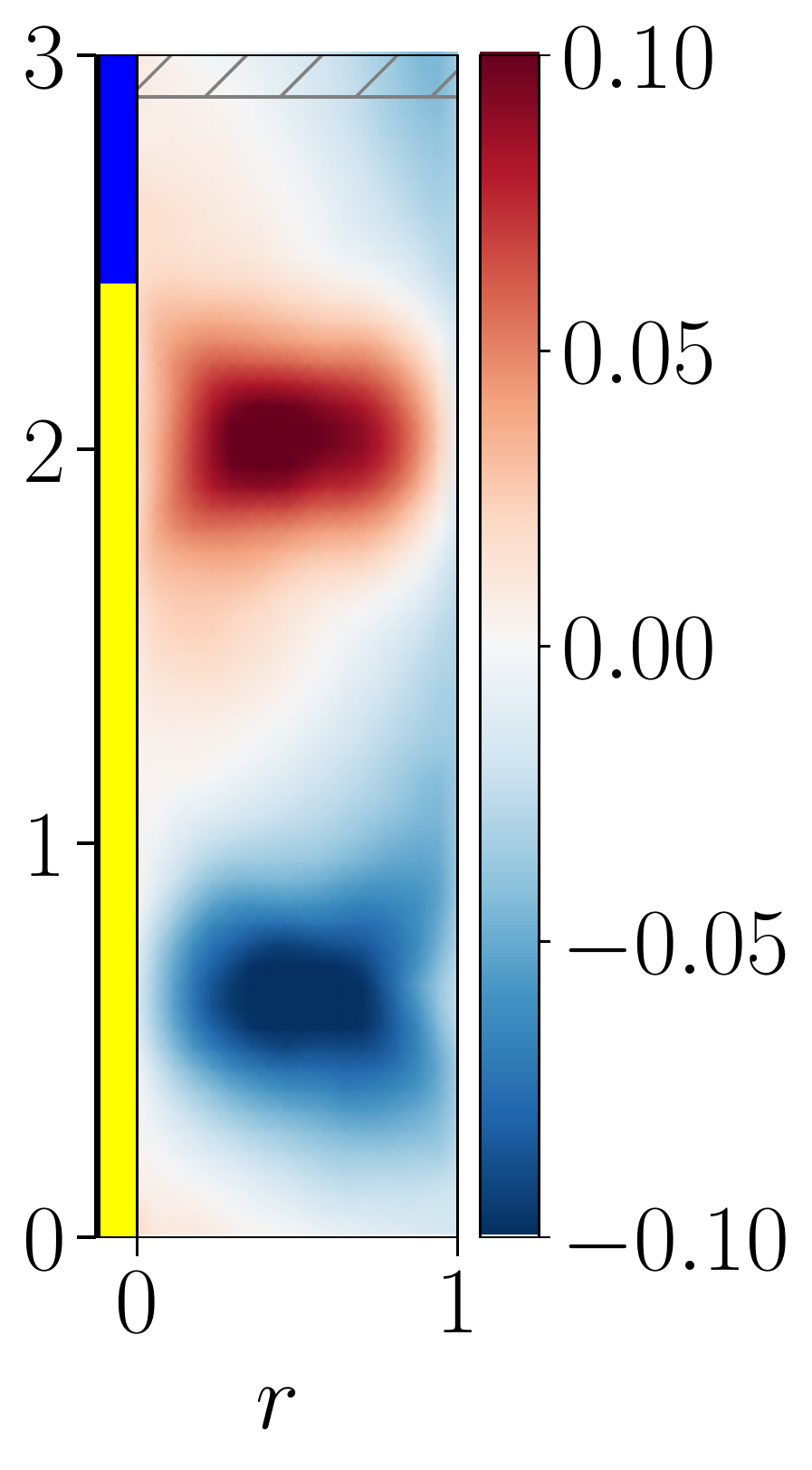} \\
  \includegraphics[height=0.3\linewidth]{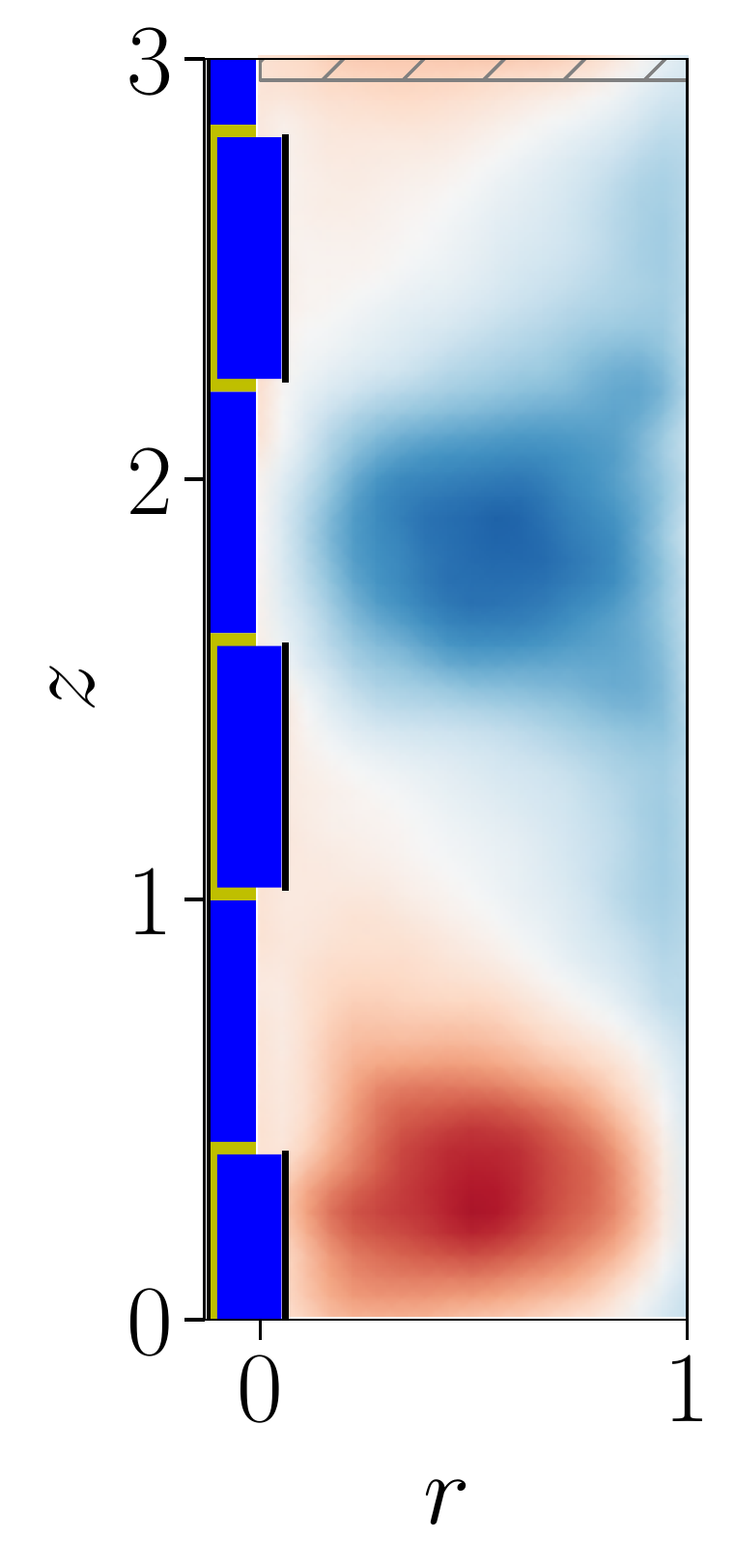}
  \includegraphics[height=0.3\linewidth]{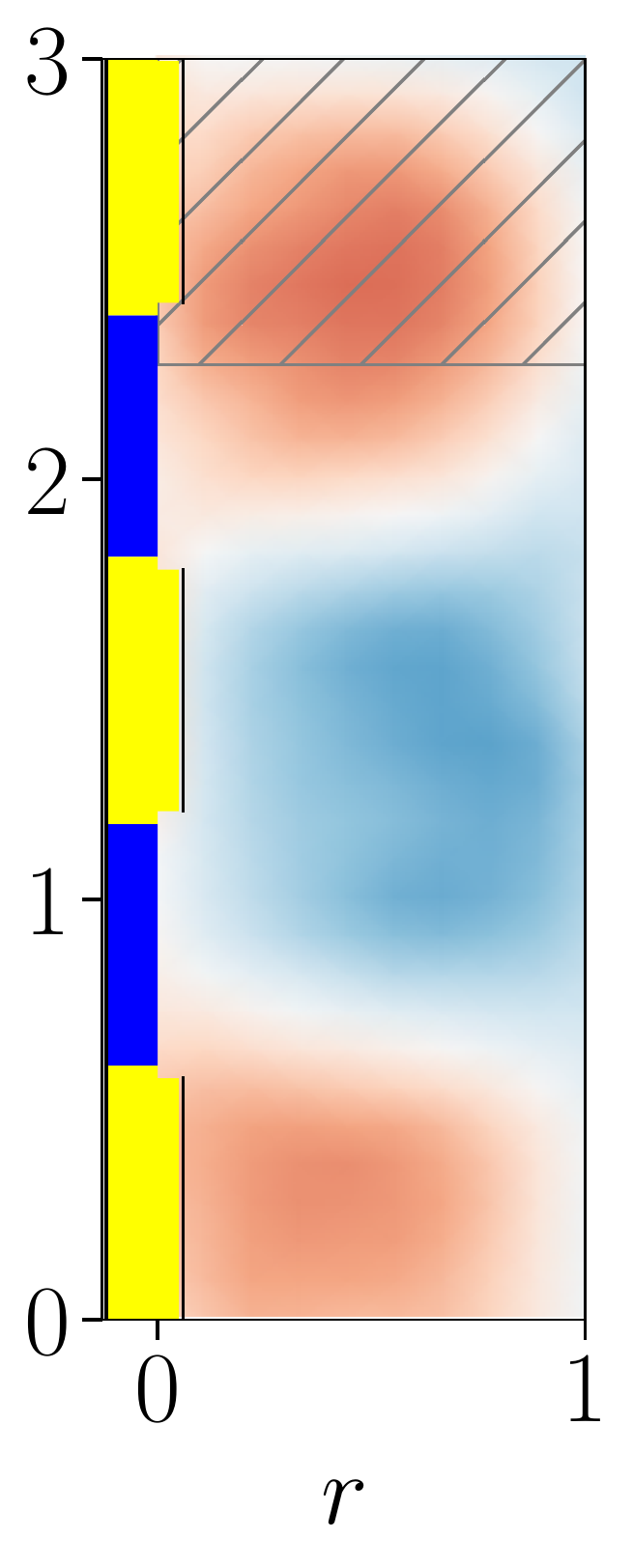}
 \includegraphics[height=0.3\linewidth]{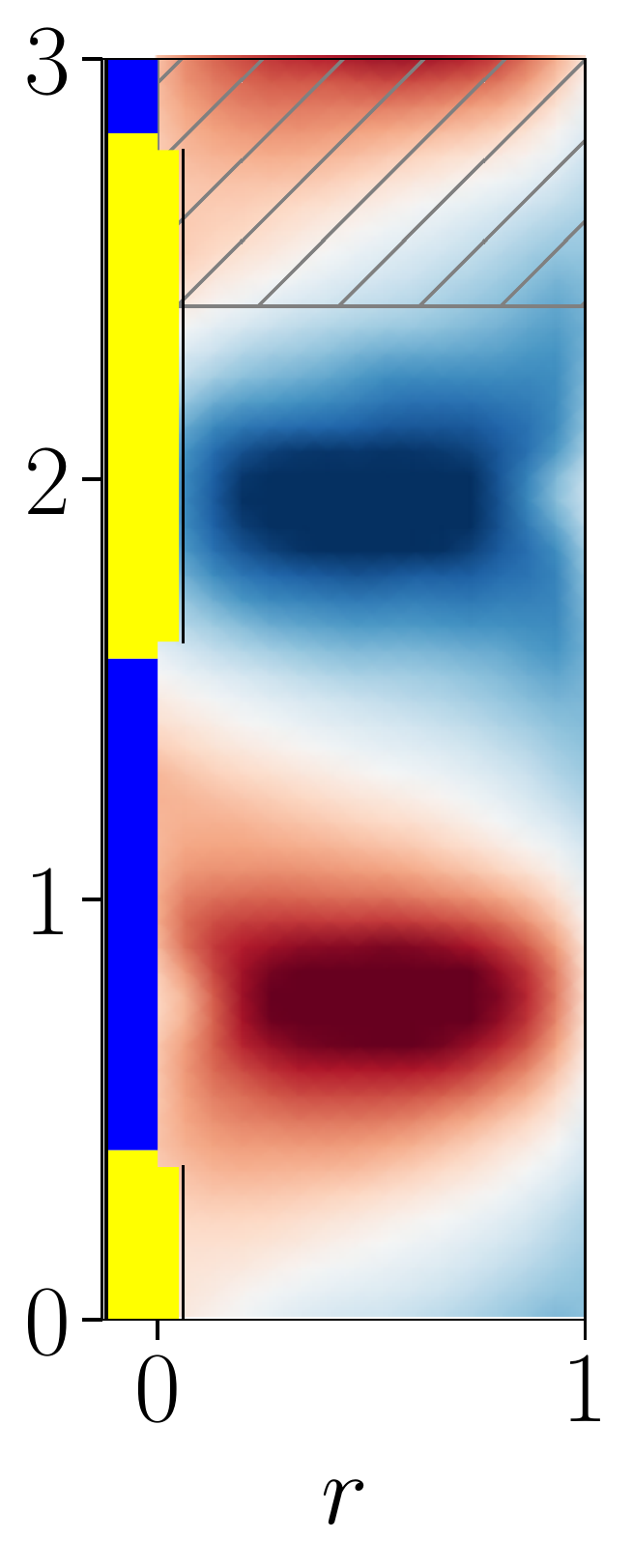}
 \includegraphics[height=0.3\linewidth]{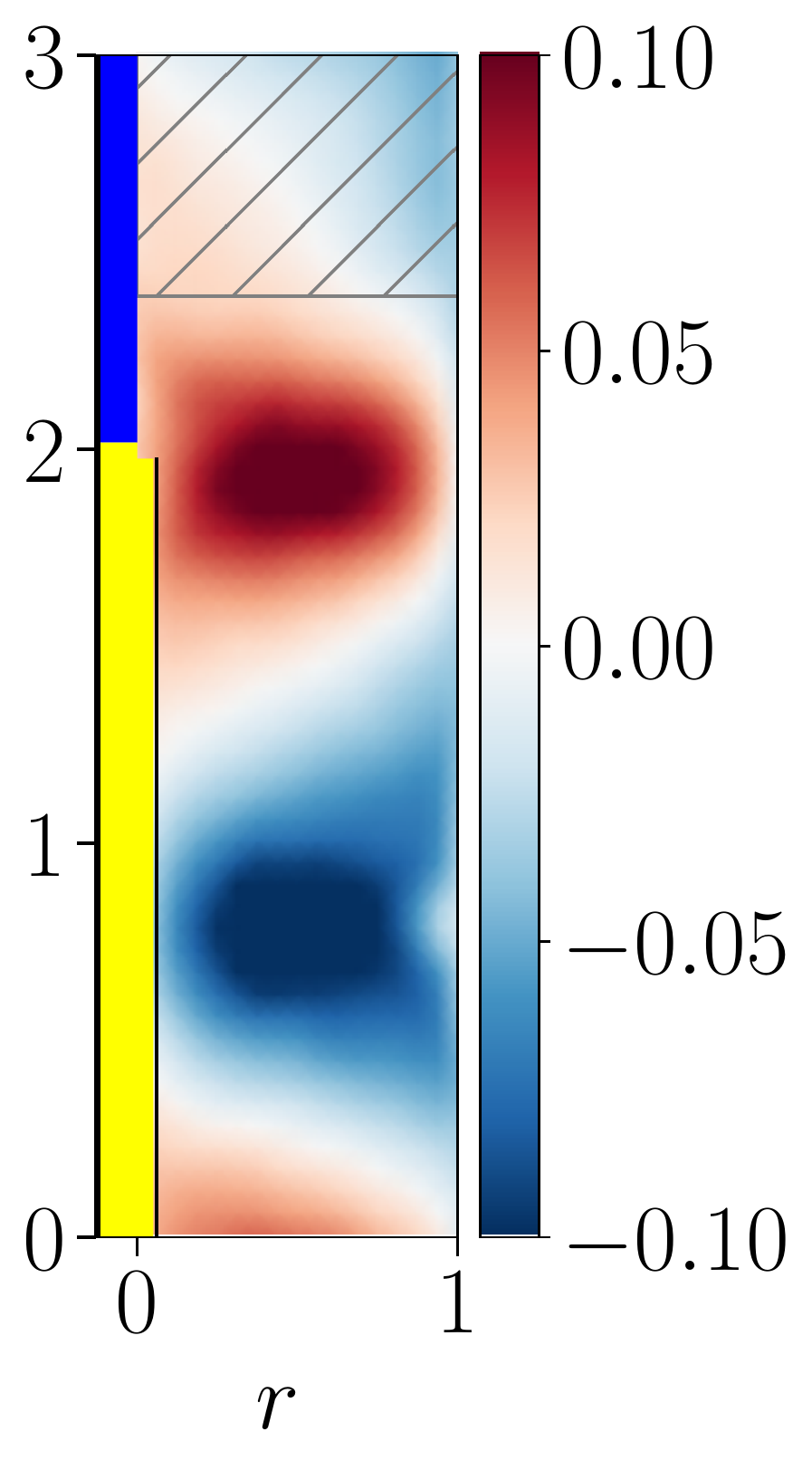} \\
\caption{Temporally-averaged radial velocity results $\langle u_r \rangle_t$ from experiments at $Re_i=10^4$. Top row, left to right: no-slip turbulent TCF; flat superhydrophobic (SHP) patterns with wavelengths $\lambda_z=1.2$,  $\lambda_z=2.4$, and $\lambda_z=4.8$. Bottom row, left to right: at  $Re_i=10^4$, stepped pattern with no SHP coating with wavelength $\lambda_z=1.2$; stepped SHP patterns with wavelengths $\lambda_z=1.2$, $\lambda_z=2.4$, and $\lambda_z=4.8$. The areas striped in gray represent the limits of one roll pair, as measured through the auto-correlation.} %Temporally-averaged radial velocity at $Re_i=2 \times 10^4$ and $\lambda_z=1.2$ for (\textbf{I}) flat and (\textbf{J}) stepped SHP patterns. Velocity plots are striped to show one pair of rolls. (\textbf{K}) Roll wavelength $\lambda_{TR}$ obtained from autocorrelations, and (\textbf{L}) roll strength $\sigma_r$ for various flat \protect\redtrinf\ and stepped \protect\bluesqnf\ SHP pattern wavelengths at $Re_i=10^4$. The black horizontal lines in \textbf{K} and \textbf{L} represent the reference no-slip roll wavelength and roll strength, respectively. (\textbf{M}) Roll strength $\sigma_r$ and (\textbf{N}) dimensionless torque $G$ for no-slip \protect\blackcirc, $\lambda_z=1.2$ flat SHP \protect\redtri\ and $\lambda_z=1.2$ stepped SHP \protect\bluesq\ patterns, at different $Re_i$. The error bars in \textbf{M} represent the standard error of the mean of the torque collected from the sensor for 60,000 eddy turnover times, which corresponds to 10 minutes.}
  \label{fig:exp_vs}
\end{figure*}

In addition to the changes in roll size, we can also observe our intended effect: Taylor rolls are slightly affected for $\lambda_z=1.2$ as seen in \Cref{fig:exp_vs}B, especially when compared to other pattern wavelengths (Figs.~\ref{fig:exp_vs}C and D) and no-slip TCF (\Cref{fig:exp_vs}A). To quantify this, we define the roll strength, $\sigma_r$, using the standard deviation of the average radial velocity $\langle u_r\rangle_{t}$ along the axial direction in the mid gap, $r=0.5$. We note that this definition is different from what is used, for example, in \cite{sacco2019dynamics}, where the roll strength is defined using the amplitude of Fourier modes. However, Fourier-based approaches are not suitable for comparisons in this study due to large deviations in the roll shape discussed later. Some care must be taken when comparing $\sigma_r$ for rolls with a different $\lambda_{TR}$, as we can expect different values of $\sigma_r$ for different sized rolls, as the roll footprint varies as the roll wavelength changes (c.f. \citep{ostilla2016effect} and \Cref{sec:expcomp}). %In particular, we can expect larger values of $\sigma_r$ for smaller rolls due to a greater variance in radial velocities being achieved in a smaller spatial extent.}

The right panel of \Cref{fig:exp_trsigma} ( \protect\redtrinf\ line ) shows the roll strength $\sigma_r$ for all treatments at $Re_i=10^4$. We first observe that the roll strength corresponding to the pattern wavelength $\lambda_z=1.2$ is lower than that of the no-slip TCF. The axial signature of the roll is weakened (as shown in \Cref{fig:exp_vs}B), and this is reflected as a smaller value of $\sigma_r$. The wavelength of this particular SHP pattern is almost equal to the size of a single large-scale structure in the flow (or half the roll wavelength). This matches our earlier numerical observations from \cite{jeganathan2021controlling}, which also showed dramatic effects on the large-scale structures of TCF when the wavelength of the ideal free-slip pattern is equal to the size of a single structure or half the roll wavelength. Physically, this weakening is caused by the axial heterogeneity, which would generate a secondary flow in a shear flow if none are present \citep{anderson2015numerical}. However, as there already is a secondary flow with a different wavelength, this heterogeneity instead induces a mismatch which destructively interferes with the existing structures, and, as a result, weakens them.

We now turn to the cases with $\lambda_z=2.4$ and $\lambda_z=4.8$. Because secondary flows have changed size in response to heterogeneity, as seen in the left panel of \Cref{fig:exp_trsigma} ( \protect\redtrinf\ line ), the variation in roll size changes other characteristics of the roll \citep{ostilla2015effects}. Hence, the variation in roll statistics, such as $\sigma_r$, can potentially be attributed to the difference in roll wavelength as well as to the effect of superhydrophobicity (c.f.~\Cref{sec:expcomp}). In the case of $\lambda_z=2.4$, the value of $\sigma_r$ is higher than the value for the no-slip case, yet $\lambda_{TR}$ is also smaller, so no conclusions can be drawn. The case with $\lambda_z=4.8$ shows a slightly lower value of $\sigma_r$ than the base-line case, but $\lambda_{TR}$ is slightly higher, again preventing us from drawing clear conclusions and distinguishing the effect of roll weakening through the SHP treatment from the variation in the roll through changing axial wavelength. In summary, while we can see that the treatment with $\lambda=1.2$ is somewhat effective, as $\sigma_r$ is lower with a smaller value of $\lambda_{TR}$, we cannot point out an optimum flat SHP wavelength as in \cite{jeganathan2021controlling}, at which the Taylor rolls could be weakened, or distinguish the effects of different $\lambda_z$. 

Furthermore, there is the possibility that the system could have solutions with different number of rolls for the same $\lambda_z$. In TC experiments, the formation of different solutions (or roll states) is achieved through control of the cylinder acceleration and phase space trajectory \citep{coles1965transition,huisman2014multiple,wen2020controlling}. In our experiment, we do not observe multiplicity of solutions in most cases: with the acceleration profile detailed in the methods section, as well as other acceleration profiles, we reliably obtain a Taylor roll wavelength of $\lambda_{TR}=2.85$ for the no-slip TC flow. On the other hand, for the special case of a $\lambda_z=1.2$ flat SHP pattern, we can achieve states with different $\lambda_{TR}$ even when the system was started up with a similar acceleration profile for the cylinders. In \Cref{fig:exp_trsigma}, we have included an additional data point (\textcolor{green}{$\rhd$} marker) that denotes another experimentally-accessible state, seen more rarely than the solution shown in \Cref{fig:exp_vs}B. This state, with a $\lambda_{TR}=2.27$ wavelength, has a larger value for $\sigma_r$ when compared to the other experimentally realizable solution, again showcasing the trend that smaller values of $\lambda_{TR}$ tend to lead to larger values of $\sigma_r$. This means that to fairly assess the treatment, one must fix the roll size, such that the roll modification is not simply a matter of the system finding it easier to access different solutions when the flat SHP pattern is present. Furthermore, to make the treatment weaken the roll in a reproducible manner, one must fix the roll size which can be a challenging task.

Now, we turn to the stepped SHP patterns. Since the stepped SHP pattern is formed due to a combination of the step feature caused by the abrasive adhesive tape and the SHP coating, it is important to assess whether the steps themselves affect the flow. To show the effect of steps, we have used uncoated smooth filler tape whose thickness the same height as the stepped SHP surface and pattern wavelength of $\lambda_z=1.2$ and presented the temporally averaged velocity in \Cref{fig:exp_vs}E for $Re_i=10^4$. Although there is some disturbance caused by the steps in the flow, arge-scale structures which can be identified as Taylor rolls still remain. The data is represented as \textcolor{magenta}{$\circ$} in \Cref{fig:exp_trsigma}. The roll wavelength is $\lambda_{TR}=2.95$, very different from the applied $\lambda_z$, and larger than the no-slip wavelength. The associated value of $\sigma_r$ is smaller than that of the no-slip case. However, due to the increased roll wavelength, it cannot be linked to a weakening of the roll.

Having checked this, Figs.~\ref{fig:exp_vs}F-H show the effect of different stepped SHP pattern wavelengths. First, we observe that, unlike flat SHP, the roll size is now fairly constant: $\lambda_{TR} \approx 2.3$ as shown in the left panel \Cref{fig:exp_trsigma} ( \protect\bluesqnf\ line ), as intended. We hypothesize that the combination of steps with SHP coating reduces the number of possible solutions and helps to fix the roll size. Since the rolls are now comparable across different pattern wavelengths, any observed change in roll strength can be attributed only to the SHP pattern inhomogeneity and not to the size of the formed roll. Furthermore, fixing the roll size increases the chances of affecting them by using precise SHP pattern wavelengths determined by theoretical and numerical methods. The success of this approach is evident from looking at the resulting velocity field for the stepped SHP pattern wavelength of $\lambda_z=1.2$ in \Cref{fig:exp_vs}F. We again quantify this effect using the roll strength $\sigma_r$, and show the results in the right panel of \Cref{fig:exp_trsigma} ( \protect\bluesqnf\ line ). Unlike the flat SHP pattern, we observe a distinct trend of increasing the roll strength with pattern wavelength due to the fixed roll size in the system. We also note the roll strength is the lowest for the stepped SHP $\lambda_z=1.2$ case, which corresponds to the heavily-altered roll state observed in \Cref{fig:exp_vs}F. Therefore, once the roll sizes are fixed, the SHP pattern with $\lambda_z=1.2$ is revealed to substantially weaken Taylor rolls, which corresponds to the heterogeneity wavelengths of about half the size of the Taylor roll, as hypothesized in \cite{jeganathan2021controlling}. We note that when applying the stepped coating, we reliably obtain the same roll size in the experiments, unlike for the flat treatment, ensuring the reproducibility of the roll modification and that the variations in $\sigma_r$ can be linked to a roll weakening. 

We emphasize that these experiments show that theoretical results can be achieved in a real-world laboratory setting using commercially available treatments. This is further demonstrated by Movie M2 of Appendix \ref{sec:vid}. This result should, however, not be taken to mean that a-priori, $\lambda_z=1.2$ is better than say $1.3$ or $1.4$ in disrupting the existing structures, but that treatments with a wavelength equal to approximately half the natural size of the rolls disrupt the rolls better than those with larger wavelengths. 

\begin{figure*}
\centering
~\hspace{1cm}
 \includegraphics[height=0.3\linewidth]{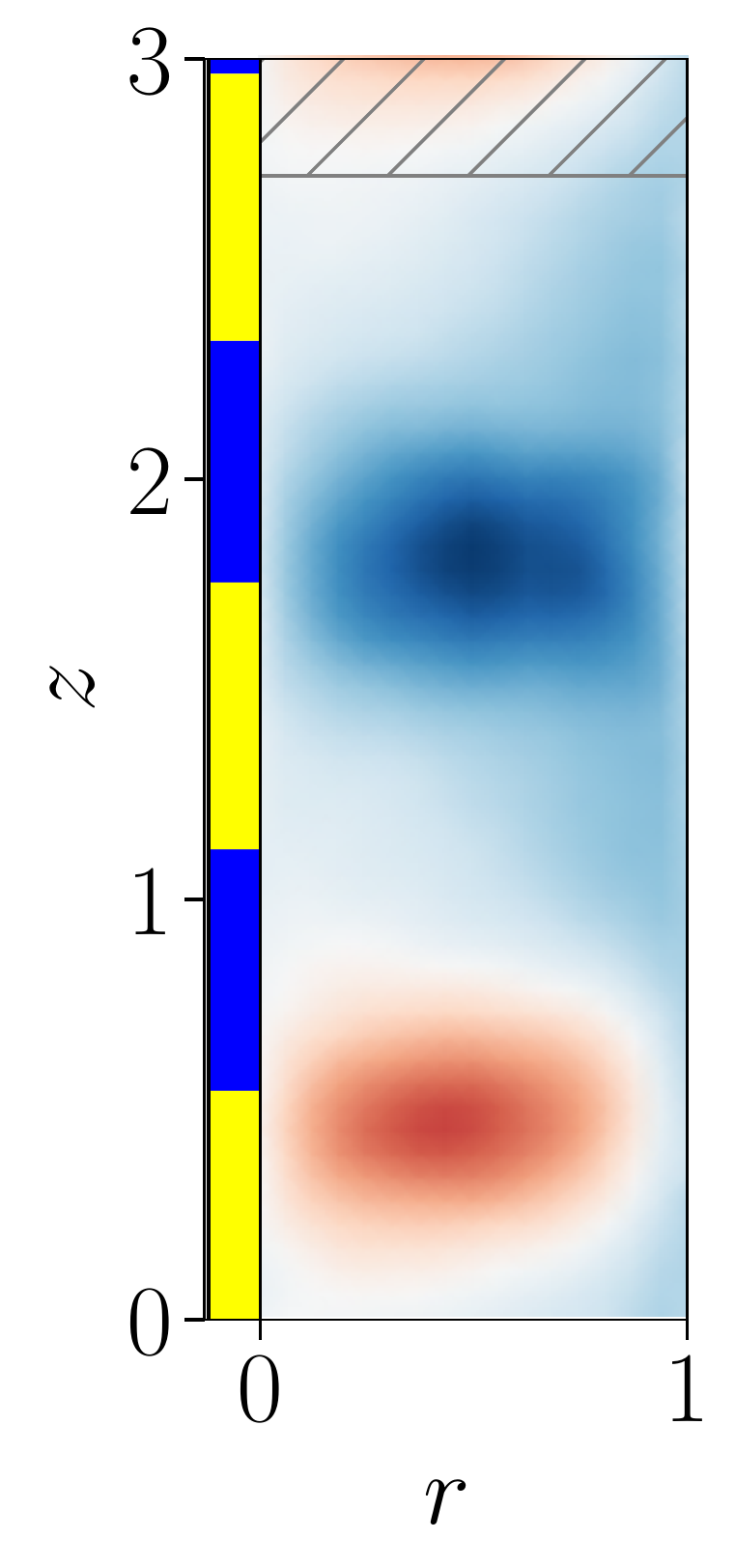}
 ~\hspace{0.3cm}
 \includegraphics[height=0.3\linewidth]{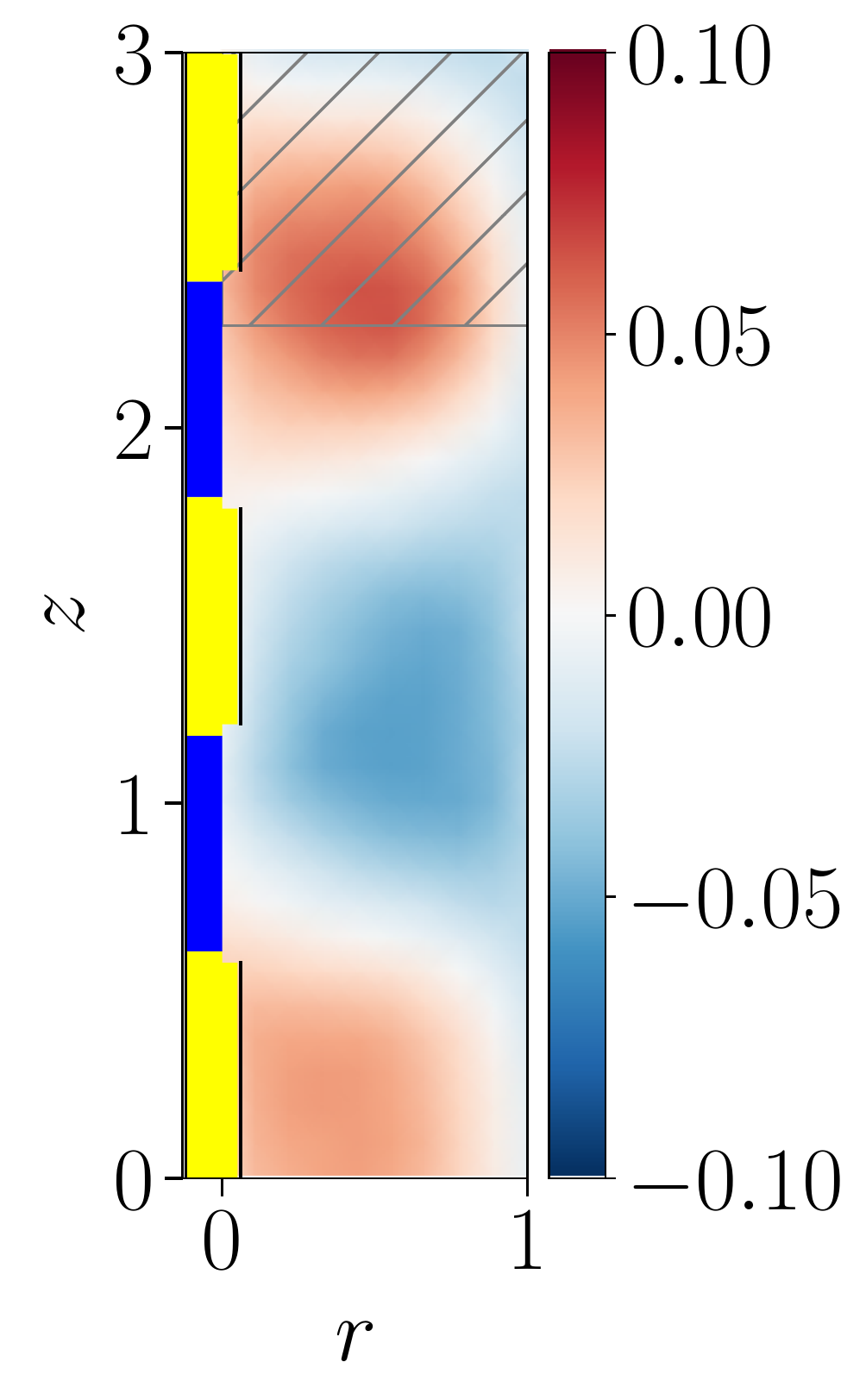}
 \hspace{0.3cm}
 \includegraphics[width=0.45\linewidth]{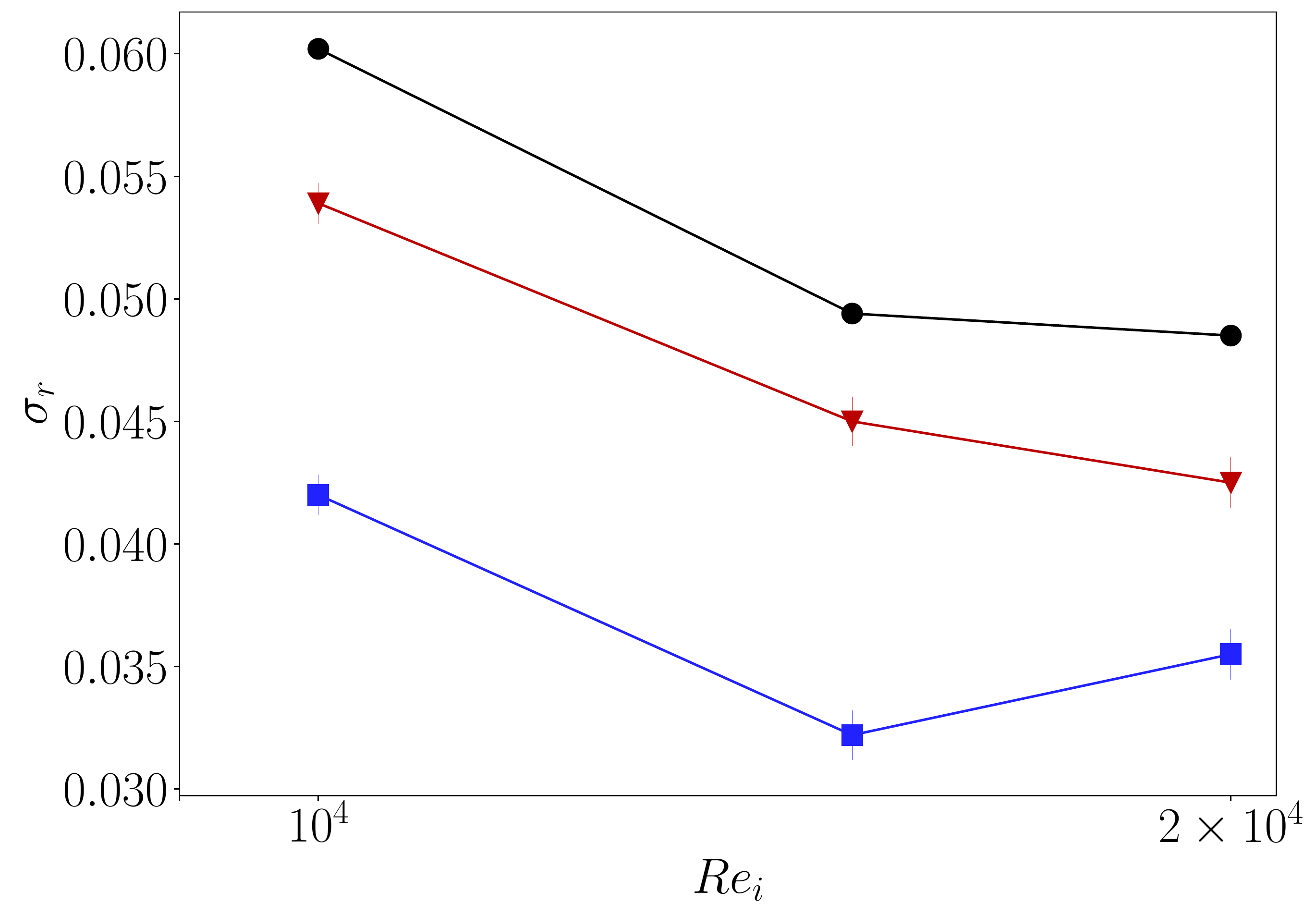}
  \includegraphics[width=0.45\linewidth]{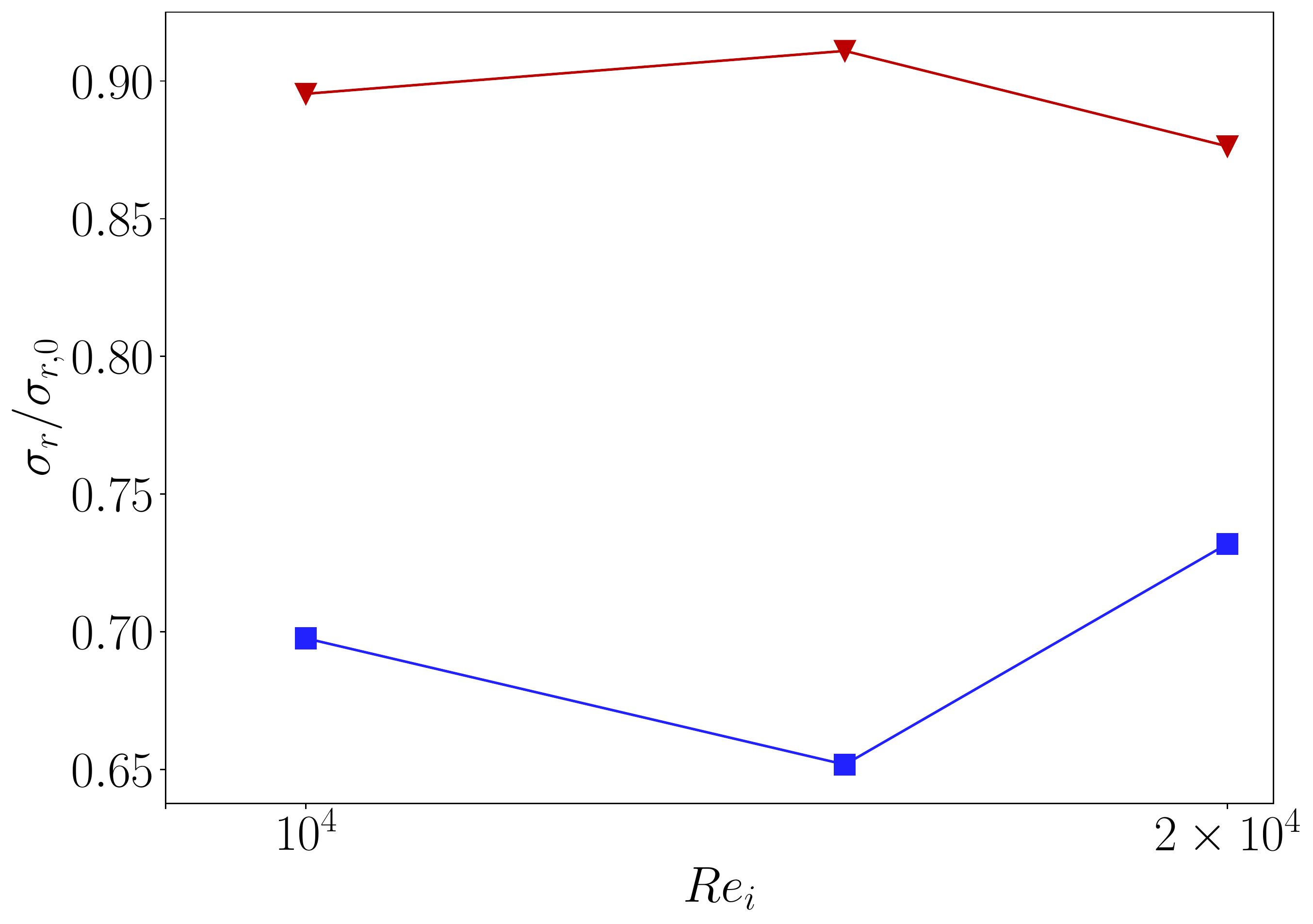}
  \includegraphics[width=0.45\linewidth]{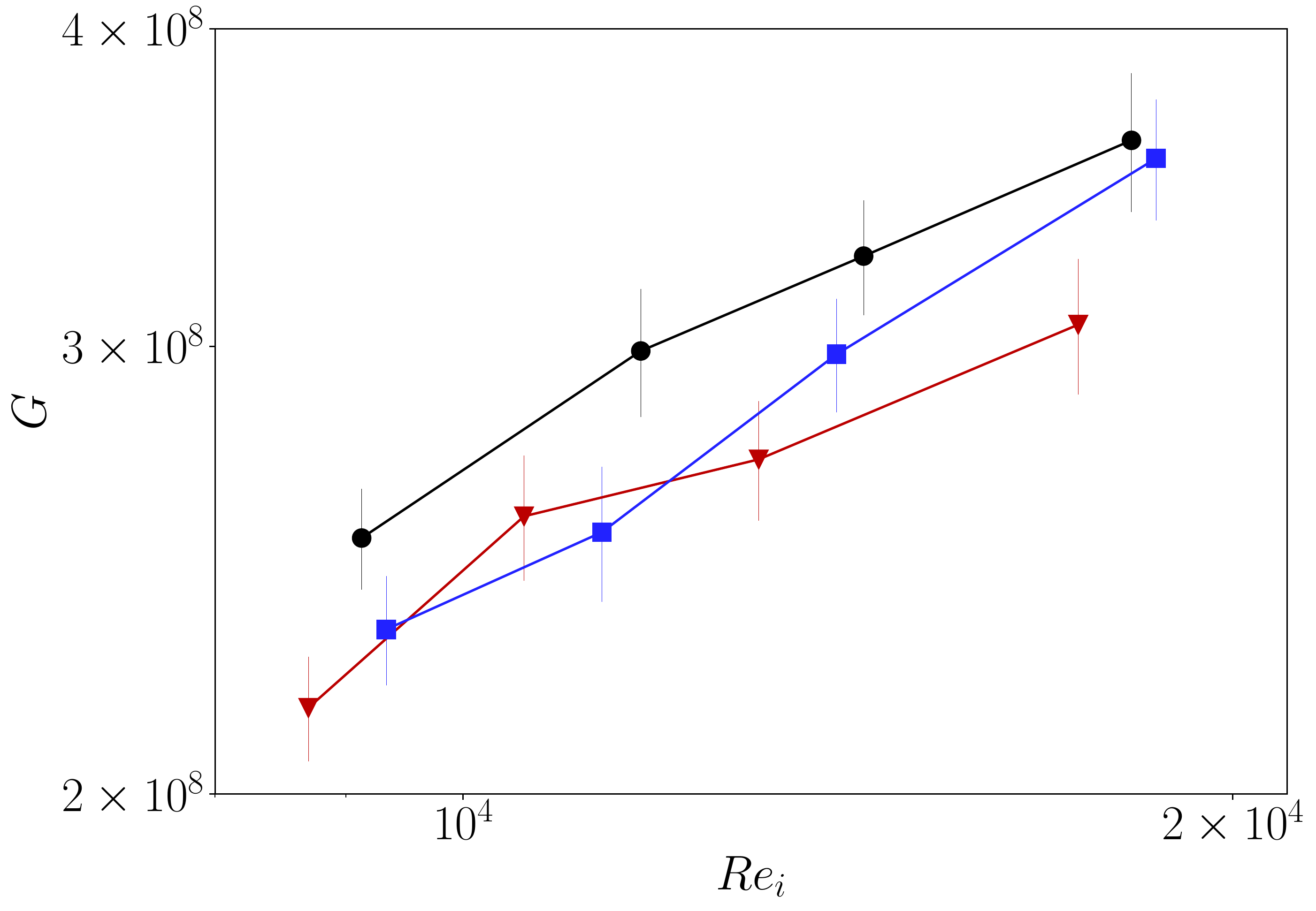}
\caption{ Reynolds number dependence. Top row, left and middle: Temporally averaged radial velocity at $Re_i=2\times 10^4$ for flat SHP (left) and stepped SHP (middle) treatments with $\lambda_z=1.2$. Right: Roll strength $\sigma_r$ for no-slip \protect\blackcirc, $\lambda_z=1.2$ flat SHP \protect\redtri\ and $\lambda_z=1.2$ stepped SHP \protect\bluesq\ patterns, at different $Re_i$. Bottom row, left: roll strength from treated cases divided by the roll strength for the no-slip reference case. Reft: Dimensionless torque $G$ for the same cases. The error bars represent the standard error of the mean of the torque collected from the sensor for 60,000 eddy turnover times, which corresponds to 10 minutes.}
  \label{fig:exp_vsre}
\end{figure*}

We have checked the effect of increasing Reynolds numbers (and of larger physical shear experiments on our SHP treatment) by extending the analysis to a larger $Re_i= 2\times 10^4$. We visually show the weakened rolls corresponding to the best treatment wavelength of $\lambda_z=1.2$ for both the flat and the stepped SHP patterns at $Re_i=2 \times 10^4$ in Figs. \ref{fig:exp_vsre}A and B respectively. The results show slight change compared to Figs. \ref{fig:exp_vs}B and F, corresponding to $Re_i=10^4$. In addition, 
\Cref{fig:exp_vsre}C shows the roll strength, $\sigma_r$, as a function of $Re_i$ for the optimum pattern wavelength of $\lambda_z=1.2$, providing a comparison between the no-slip TCF (\protect\blackcirc\ line), flat SHP (\protect\redtri\ line), as well as stepped SHP surface (\protect\bluesq\ line). We clearly see that both the flat and the stepped $\lambda_z=1.2$ SHP patterns make Taylor rolls weaker across all $Re_i$. As expected, the stepped $\lambda_z=1.2$ SHP pattern is better at weakening the rolls in the $Re_i$ range studied compared to the flat pattern due to its ability to fix the roll size. This is further corroborated in the bottom left panel of \Cref{fig:exp_vsre}, which shows the normalized roll strength $\sigma_r/\sigma_{r,0}$, where $\sigma_{r,0}$ is the reference from the no-slip data. It can be appreciated that the flat SHP case only causes a $10\%$ reduction in $\sigma_r$, while the stepped case results in a reduction of between $25-35\%$.

Finally, we compare the effects of the best treatment wavelength $\lambda_z=1.2$ for flat and stepped SHP patterns on torque in the bottom right panel of \Cref{fig:exp_vsre}. The mean torque is displayed as a dimensionless parameter, $G=\langle \hat T \rangle_t/(\hat \rho \hat l \hat \nu^2)$, where $\langle \hat T \rangle_t$ is the time average of the discrete torque values, $\hat T$, measured by the torque sensor for 60,000 eddy turnover times (10 minutes in physical time), here $\hat l$ is the length of the inner cylinder, and $\hat \rho$ and $\hat \nu$ are the density and kinematic viscosity of the working fluid respectively. The uncertainty in the mean torque is quantified through the standard error of mean \citep{gul2018experimental}. The standard error of the mean is given by $\hat\epsilon_T=\hat\sigma_T/\sqrt{N}$, where $\hat\sigma_T$ is the standard deviation of the torque and $N$ is the number of torque samples collected. For both the flat and stepped SHP patterns, we notice that the mean torque on the inner cylinder is lower when compared to the regular no-slip TCF. This is a clear indication of the drag-reduction property of the SHP surfaces. They show relatively similar scaling laws $G \sim Re_i^\alpha$, even if the torque reductions are smaller than those hypothesized in \cite{jeganathan2021controlling}, a point to which we will return below. We also note that the weakening of the rolls, as quantified through $\sigma_r$, does not seem to be a good predictor of the torque decrease for the flat SHP case. While this is unlike what was seen in the simulations \citep{jeganathan2021controlling}, we again note that the experiments with flat SHP patterns tend to actually achieve a multiplicity of possible solutions, and this could be causing the erratic increases of $G$. 

\section {Direct Numerical Simulations}
\subsection{Numerical Methods}
To further assess the potential applicability of SHP coatings, we have performed a series of Direct Numerical Simulations (DNS) of a similar TCF system using a second-order energy-conserving finite-difference code regularly used in our research group to simulate such systems \citep{van2015pencil,jeganathan2021controlling}. DNS of TCF are performed in a rotating frame of reference by solving the non-dimensional incompressible Navier-Stokes equations:

%%% Numbered equation
\begin{align}\label{1.1}
\frac{\partial\textbf{u}}{\partial t} + \textbf{u}\cdot\nabla\textbf{u} + R_\Omega (\textbf{e}_z \times \textbf{u})= -\nabla p + Re_s^{-1} \nabla^2\textbf{u},
\end{align}

\noindent with the incompressibility condition

\begin{align}\label{1.2}
\nabla\cdot\textbf{u}=0,
\end{align}

\noindent where \textbf{u} and $p$ are the non-dimensional velocity and pressure respectively; t is the non-dimensional time; $Re_s$ is the shear Reynolds number defined below; $\textbf{e}_z$ is the unit vector in the axial direction, and $R_\Omega$ the Coriolis parameter defined below.

The rotating frame is chosen such that the velocities of both cylinders are equal and opposite, $\pm \hat U/2$ in dimensional terms. The equations are non-dimensionalized using this velocity $\hat U$ and the gap width $\hat d$. This results in two non-dimensional control parameters, the shear Reynolds number $Re_s=\hat U \hat d/\hat \nu$ and the Coriolis parameter $R_\Omega=2 \hat \Omega \hat d/\hat U$, where $\hat \nu$ is the kinematic viscosity of the fluid and $\hat \Omega$ is the dimensional rotational velocity of the rotating frame.

The domain is taken to be axially periodic, with a periodicity length $\hat L_z$, which can be expressed non-dimensionally as an aspect ratio $\Gamma_z=\hat L_z/\hat d$. Following Ref.~\cite{jeganathan2021controlling}, the domain is set to be axially periodic with a dimensionless axial periodicity length of $\Gamma_z=\hat{L}_z/\hat{d}=2.33$. This fixes the wavelength of the roll pair $\lambda_{TR}=\Gamma_z$ and forces the domain to contain a single roll pair. The radius ratio, $\eta=\hat r_i/\hat r_o$ is fixed to $\eta=0.83$, where $\hat r_i$ and $\hat r_o$ are the radius of the inner and outer cylinders, corresponding to the experiments. We also impose a rotational symmetry of order $n_{sym}=10$, corresponding to streamwise periodicity length of around $2\pi$ half-gaps, large enough to obtain asymptotic torque and mean flow statistics \citep{ostilla2015effects,jeganathan2021controlling}.

Spatial discretization is performed using a second-order energy-conserving centered finite-difference scheme. Time is advanced using a low-storage third-order Runge-Kutta for the explicit terms and a second-order Crank-Nicholson scheme for the implicit treatment of the wall-normal viscous terms. More details of the algorithm can be found in previous studies \cite{verzicco1996finite, van2015pencil}. The code has been heavily validated for the TCF problem \citep{ostilla2014exploring}. The spatial resolution used is $n_\theta\times n_r\times n_z = 384\times 512\times 768$ in the azimuthal, radial, and axial directions, respectively, following \cite{sacco2019dynamics}.

In a classical TCF problem, the cylinders have a no-slip boundary condition, where the velocity of the fluid at the wall matches the velocity of the cylinder. However, in the present study, we alternate no-slip and finite-slip boundary conditions at the wall. Finite-slip boundary conditions are expressed by the combination of (1) a no penetration ($u_r=0$), and (2) the condition that the two velocity components tangential to the wall equal the slip length times their respective normal derivatives. In non-dimensional terms, this is expressed as $u_{\theta}= b \partial_r u_\theta$ and $u_z= b \partial_r u_z$. We implement the finite-slip boundary condition by modifying the shear stress $\tau$ originating from the wall at the first point on the grid. This is done by modifying the viscous term, which is first approximated using a finite difference of shear stresses ($[\tau^+-\tau^-]/\Delta r$). Then, these shears are approximated using a finite difference of velocities, $u_z$ or $u_\theta$, depending on the direction being considered. With some rearrangement, this results in a simple correction factor to the geometric factors which multiply the velocity difference. For example, the radial shear stress for $u_z$ at the first grid point is expressed with the following equation:

\begin{equation}
    \tau_{z,1}^{-} = Re_s^{-1} \left . \displaystyle\frac{\partial u_{z}}{\partial r}\right |_{r_1} = Re_s^{-1} \displaystyle\frac{u_{z}(r_1) - u_{z}(r_i) }{r_1-r_i},
    \label{eq:finiteslip}
\end{equation}

\noindent where $u_{z}(r_i)$ is the axial velocity at the inner cylinder, $u_{z}(r_1)$ is the axial velocity at the first grid point, and $r_1$ the radial coordinate of the first grid point. In the case of no-slip, $u_{z}(r_i)$ is equal to the wall velocity (zero for the axial component), while in the case of finite-slip, $u_{z}(r_i)$ is equal to the slip velocity $u_{z,s}$. The slip velocity can be re-written as $u_{z,s}=b \partial_r u_z(r_1) = b Re_s \tau^-_{z,1}$. Expressed this way, it can be substituted back into Eq.~\ref{eq:finiteslip} and the equation is now closed. The finite difference approach of the code allows us to quickly change between no-slip, finite-slip and free-slip conditions by modifying the metric terms multiplying $\tau^-_{z,1}$, and can allow for potential extensions of this work which consider a spatially- or temporally-dependent slip-length.

The alternating no-slip and finite-slip boundaries applied in the code have a pattern wavelength of $\lambda_z=\lambda_{TR}/2=1.17\approx 1.2$, similar to experiments. We simulate pure inner cylinder rotation with an inner cylinder Reynolds number of $Re_i=10^4$ to match the experiments by setting $R_\Omega=(1-\eta)=0.17$ and $Re_s=2/(1+\eta) Re_i = 1.09\times10^4$. We then vary the dimensionless slip length, $b=\hat b / \hat d$, to determine the minimum slip length required for our treatments to be effective in weakening the secondary flows. The simulations are initialized from a zero-velocity condition, and, after a transient which usually takes around $200$ large-eddy turnover times ($\hat{d}/\hat{U}_i$) flow statistics are taken for around $600$ large-eddy turnover times. This criteria ensures that the time-averaged torque at both cylinders is equal to within $1\%$.

\begin{figure}
\centering
\includegraphics[height=0.25\textwidth]{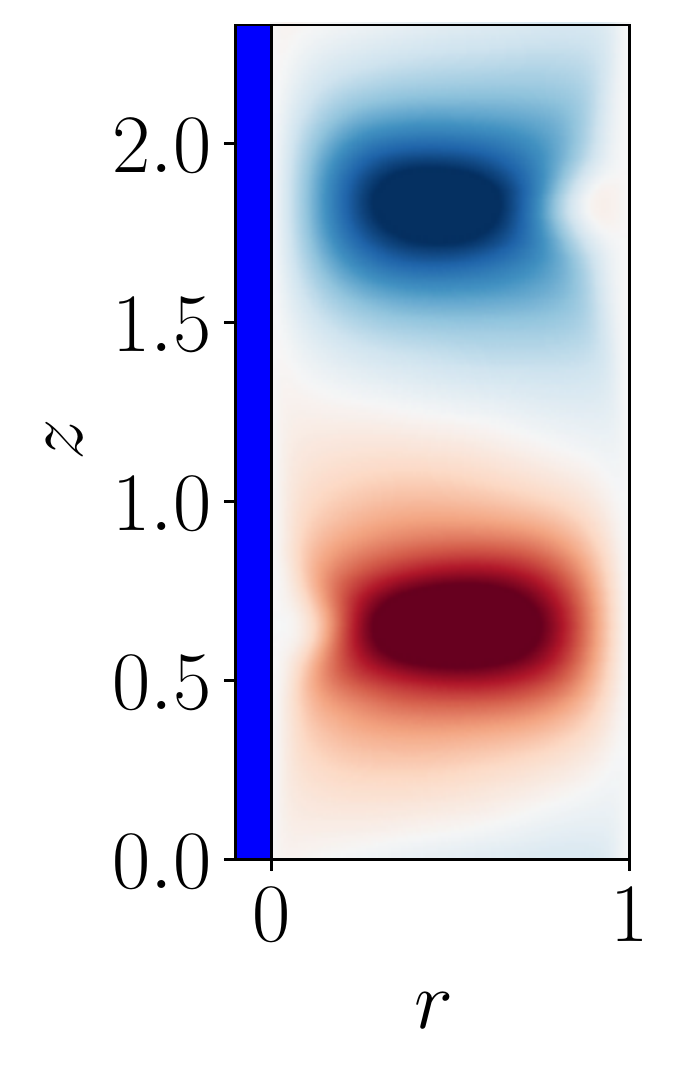}
\includegraphics[height=0.25\textwidth]{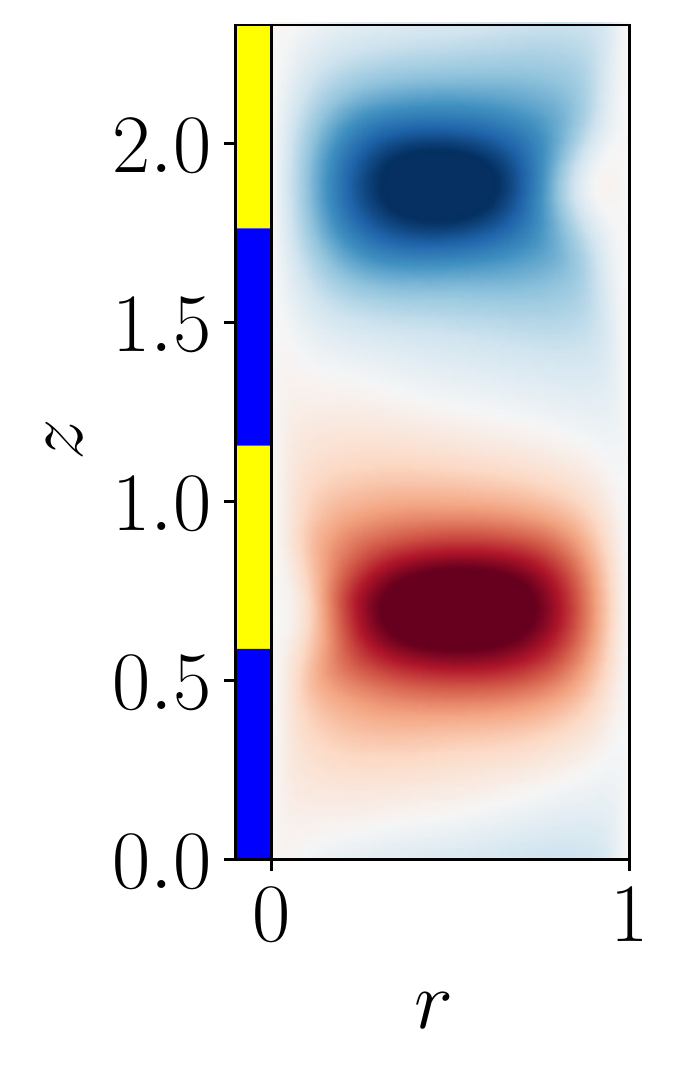}
\includegraphics[height=0.25\textwidth]{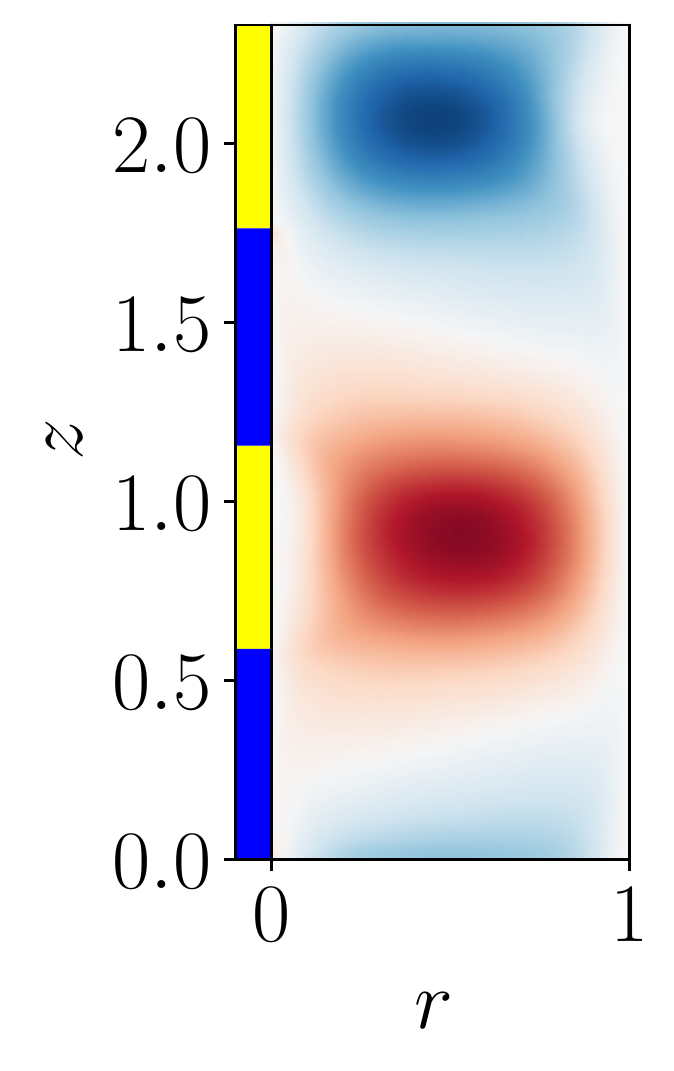}
\includegraphics[height=0.25\textwidth]{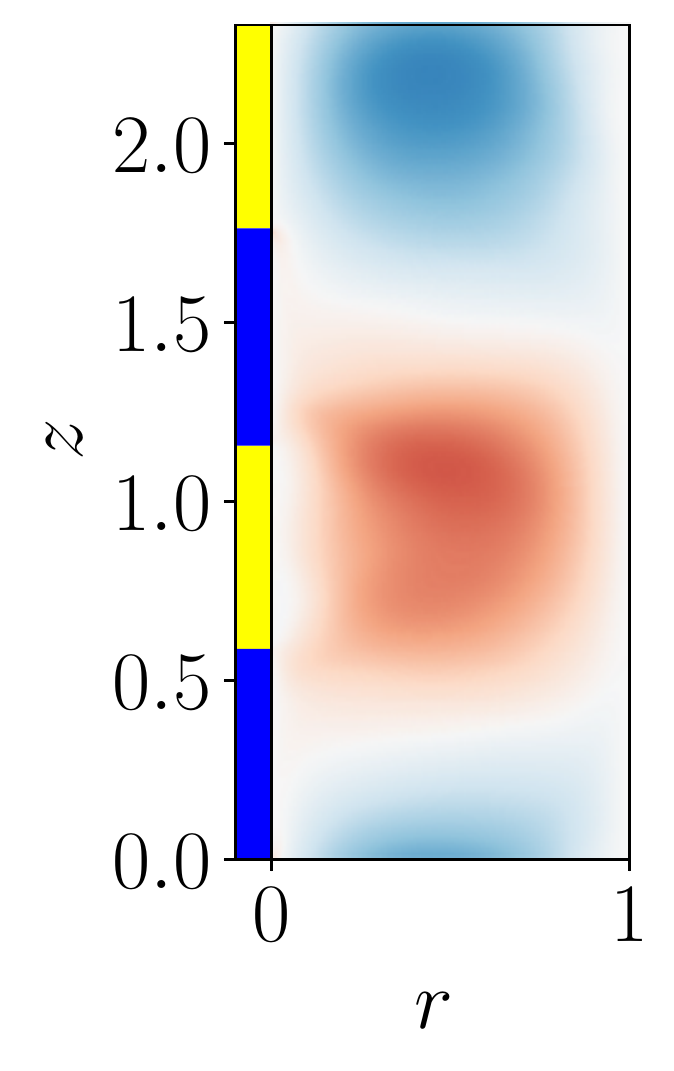}
\includegraphics[height=0.254\textwidth]{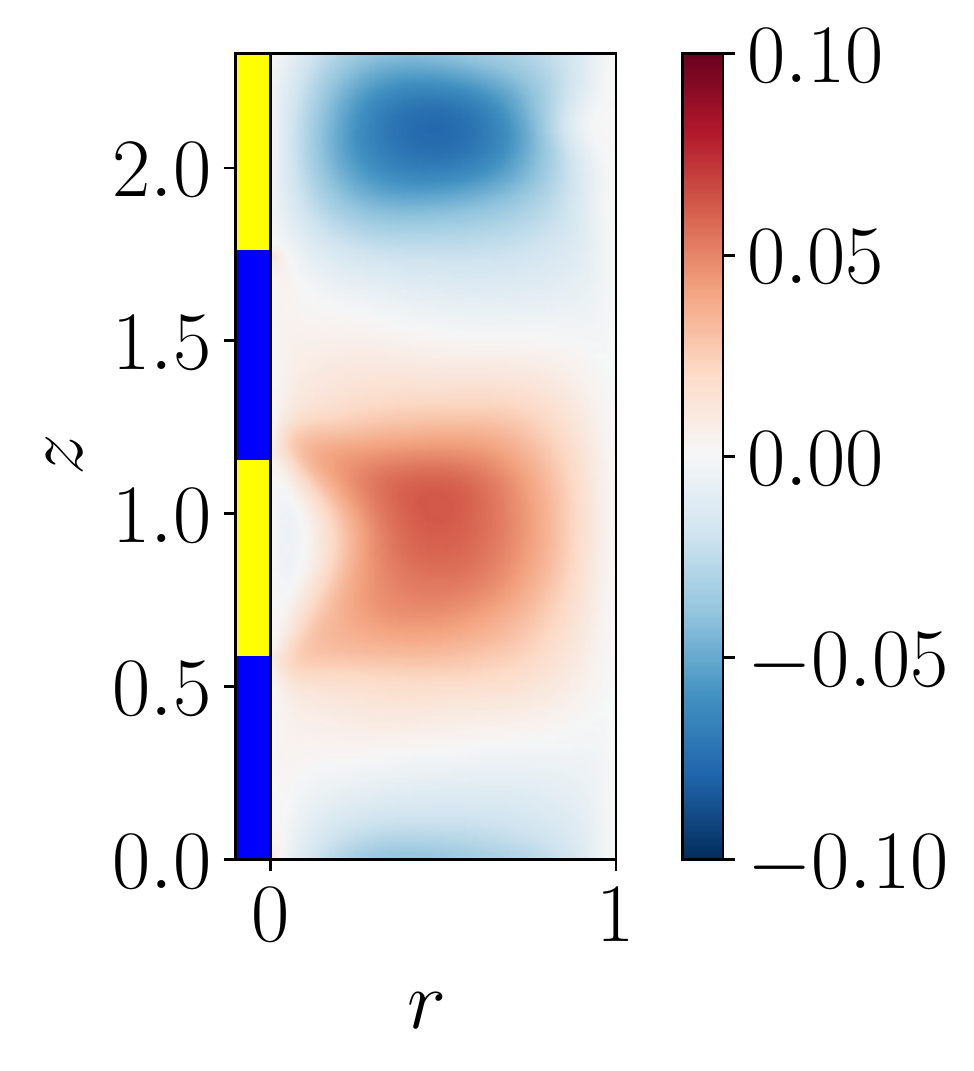}
\caption{Simulation results. Temporally and azimuthally-averaged radial velocity of flat superhydrophobic (SHP) patterns at $Re_i =10^4$, with pattern wavelength $\lambda_z=1.2$ and various dimensionless slip lengths $b=\hat b/\hat d$. From left to right: no slip ($b=0$), $b=10^{-4}$, $b=10^{-2}$, $b=1$, and free slip ($b\to\infty)$.}
%(\textbf{F}) Roll strength $\sigma_r$ and (\textbf{G}) torque $G$ for SHP $\lambda_z=1.2$ pattern at $Re_i =10^4$ for various dimensionless slip lengths $b$ and $b^+$ \protect\greendia, normalized using the reference no-slip case. Horizontal dashed lines represent the free-slip limit (perfect hydrophobicity). The green curves are the sigmoid fits shown in \Cref{eqn:norm_amp,eqn:norm_torq}. The normalized roll strength and normalized torque of the flat \protect\redtri\ and the stepped \protect\bluesq\ SHP coated TCF experiments are also shown for $Re_i =10^4$ and $\lambda_z=1.2$. The error bars in \textbf{G} represent the standard error of the mean torque collected from the sensor.}
  \label{fig:sim_vs}
\end{figure}

\subsection{Numerical Results}
To study the Taylor rolls in the DNS, we average the radial velocity temporally and azimuthally, as opposed to experiments where the PIV data along the azimuth are unavailable. Averaging azimuthally is justified by the statistical homogeneity of TCF, and this allows us to reduce the required running time of the simulations to obtain adequate statistics. The flow statistics are also averaged in time, as mentioned earlier. We note that a more detailed analysis of the temporal dynamics of turbulent Taylor rolls is available in \cite{sacco2019dynamics} for the interested reader.

\begin{figure}
\centering
 \includegraphics[width=0.47\linewidth]{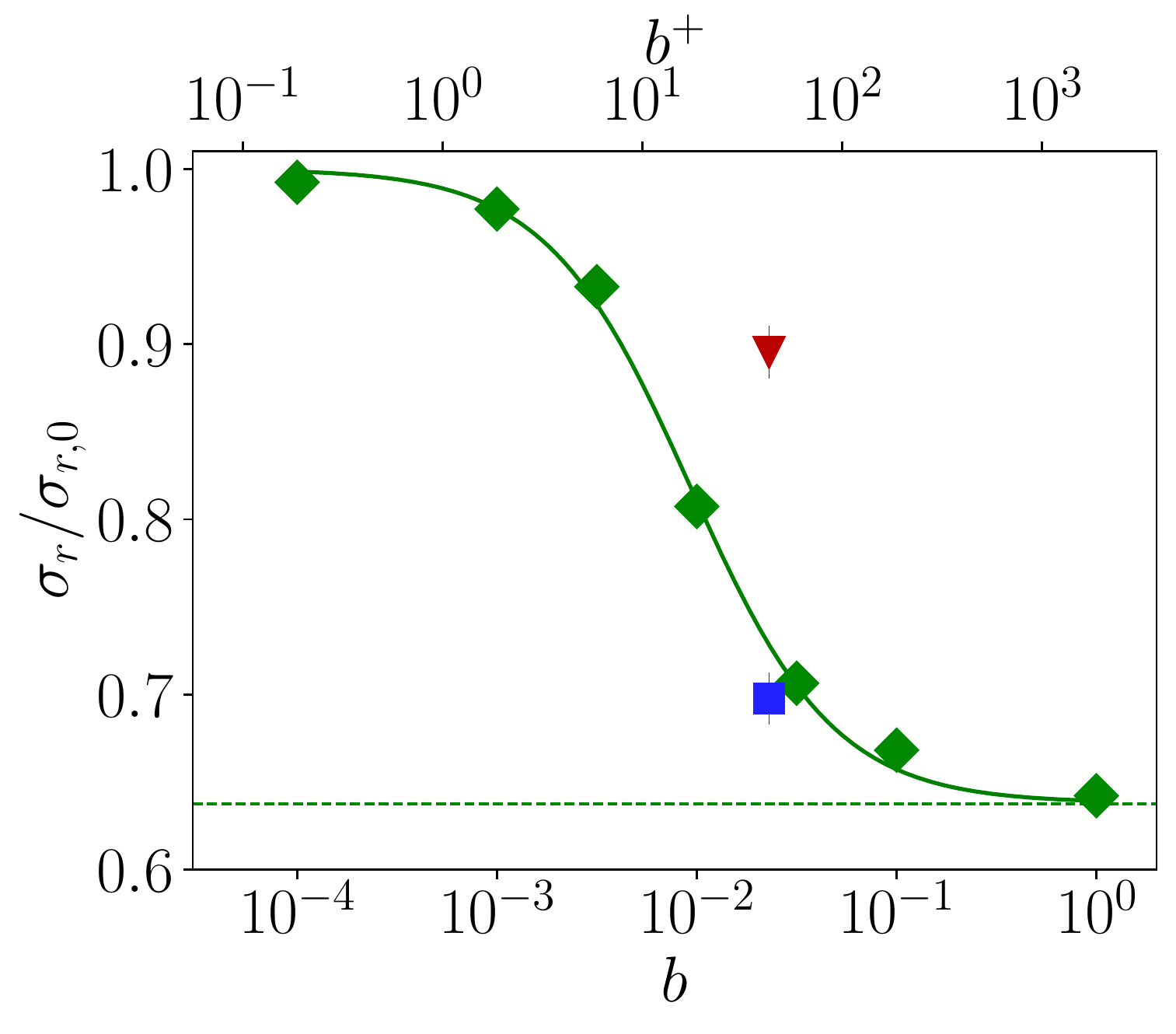}
 \includegraphics[width=0.47\linewidth]{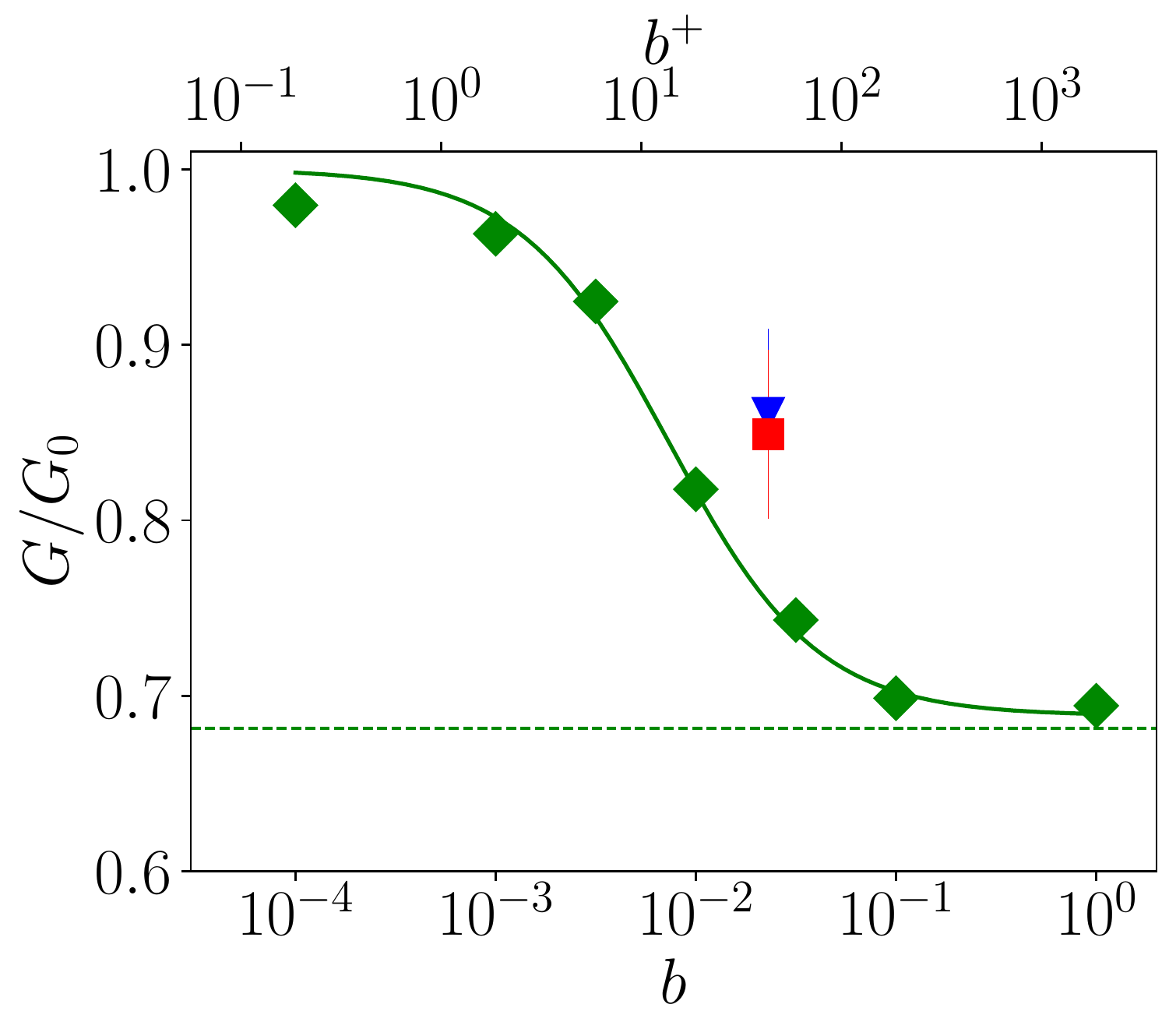}
 \caption{Roll strength $\sigma_r$ (left) and torque $G$ for SHP $\lambda_z=1.2$ pattern at $Re_i =10^4$ for various dimensionless slip lengths $b$ and $b^+$ \protect\greendia, normalized using the reference no-slip case. Horizontal dashed lines represent the free-slip limit (perfect hydrophobicity). The green curves are the sigmoid fits shown in \Cref{eqn:norm_amp,eqn:norm_torq}. The normalized roll strength and normalized torque of the flat \protect\redtri\ and the stepped \protect\bluesq\ SHP coated TCF experiments are also shown for $Re_i =10^4$ and $\lambda_z=1.2$. The error bars on the right panel represent the standard error of the mean torque collected from the sensor.}
  \label{fig:sim_re1k}
\end{figure}

Fig.~\ref{fig:sim_vs} shows the temporally and azimuthally-averaged dimensionless radial velocity, $\langle u_r\rangle_{\theta,t}$, for different simulated slip lengths ranging from untreated/no-slip ($b = 0$; Fig.~\ref{fig:sim_vs}A) to ideal treatment/free-slip ($b\to\infty$; Fig.~\ref{fig:sim_vs}E). As expected, Taylor rolls are strongest when there is no treatment. This is also shown in Movie M3. As the slip length increases, the rolls gradually weaken, as seen in Figs.~\ref{fig:sim_vs}B-D, with the most effective treatment observed when the surface is fully free-slip, as illustrated in Fig.~\ref{fig:sim_vs}E and also in Movie M4. To further quantify this effect, we plot the normalized roll strength $\sigma_r/\sigma_{r,0}$ and the normalized torque $G/G_0$ against the slip length $b$ in Fig.~\ref{fig:sim_re1k}, respectively. Here, $\sigma_{r,0}$ and $G_0$ correspond to the roll strength and torque of the untreated no-slip reference case respectively. The experimental data for the same $Re_i$ and $\lambda_z$ are also included for completeness. Remarkable agreement is observed for $\sigma_r/\sigma_{r,0}$ between the simulations and the stepped SHP experiment, where the roll size is also fixed. For the torque, both experimental cases show less reduction than the simulation. This can likely be attributed to higher torque losses in the experiments, such as those resulting from axial end caps, which are absent from DNS.

A transition in behaviour, in which treatment begins to be effective, can be observed at $b\approx3\times10^{-3}$. We fit both the roll strength and torque of DNS using a sigmoid function, which are shown in \Cref{eqn:norm_amp,eqn:norm_torq} (solid green lines in Figs. \ref{fig:sim_re1k}). The curve fits reads as:

\begin{equation}
\frac{\sigma_r}{\sigma_{r,0}}=1-\frac{0.362}{1+\exp{-1.21(4.684+\ln b)}}, {\textmd{ and}} \label{eqn:norm_amp}
\end{equation}

\begin{equation}
\frac{G}{G_0}=1-\frac{0.3113}{1+\exp{-1.135(4.872+\ln b)}}.  \label{eqn:norm_torq}
\end{equation}

\noindent As is confirmed by the fits, both curves show an inflection point at $b\approx10^{-2}$. To better understand why this happens, we must represent the slip length in inner wall units, $b^+=\hat{b} / \hat \delta_{\nu}$ (see the upper $x$-axis in Figs.~\ref{fig:sim_re1k}A and \ref{fig:sim_re1k}B), where $\delta_\nu= \hat \nu / \hat u_\tau$ is the viscous wall unit, $\hat{u}_{\tau}=\sqrt{\hat \tau_w/\hat \rho}$ the frictional velocity and $\hat \tau_w$ the averaged shear at the cylinder wall. This (re)normalizes the slip length in terms of the relevant length scales in the boundary layer. In both the normalized roll strength and torque cases of Figs.~\ref{fig:sim_re1k}, we can distinguish three regions of behavior. In the first region, where the slip length is $b^+<1$ ($b<10^{-3}$), the flow and hence the Taylor rolls remain largely unaffected. In this regime, the effect of treatment is too weak. The slip length is smaller than the smallest physical scale present in the problem. Hence, it is largely unfelt by the fluid. As the slip length increases, we reach a second region, $1<b^+<100$ ($10^{-3}<b<10^{-1}$). In this transitional region, the treatment is sufficiently effective so that the flow begins to be affected by the boundary inhomogeneity. This, in turn causes the destructive interference effects mentioned above which progressively affect the roll strength and torque. For a treatment to be effective, it has to be strong enough to achieve slip lengths located in this region, where the slip length is comparable to the viscous length-scale in the boundary layer. For example, the SHP surface used in the TCF experiments has a slip length $b \approx 0.023$, which is large enough to reach this region and therefore can weaken Taylor rolls, as shown by Fig. \ref{fig:exp_vs}F of the experimental results and DNS simulation presented as Movie M5. In the third region, $b^+>100$ ($b>10^{-1}$), the effect of inhomogeneity reaches a saturation point, where a further increase in slip length does not significantly affect rolls. The treatment behaves as if it were ideal ($b\to\infty$), as the slip length is now comparable to the largest length-scales in the flow. This saturation is reflected in the normalized roll strength and torque that tend toward the asymptotic limit of free-slip, i.e.~$b\to\infty$ or stress-free, shown as dashed black lines in \Cref{fig:sim_re1k}A and B.

\begin{figure}
\centering
 \includegraphics[width=0.47\linewidth]{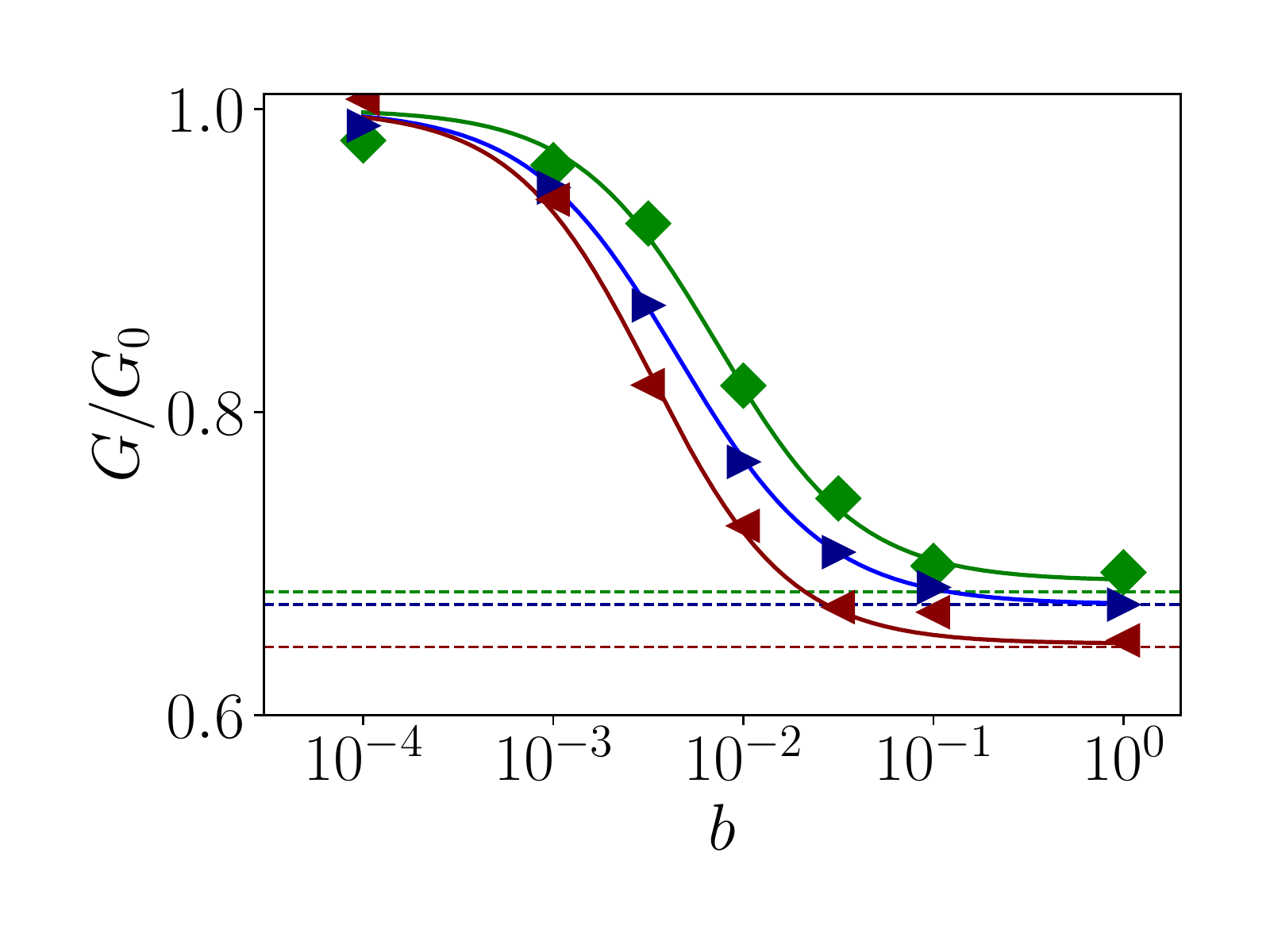}
 \includegraphics[width=0.47\linewidth]{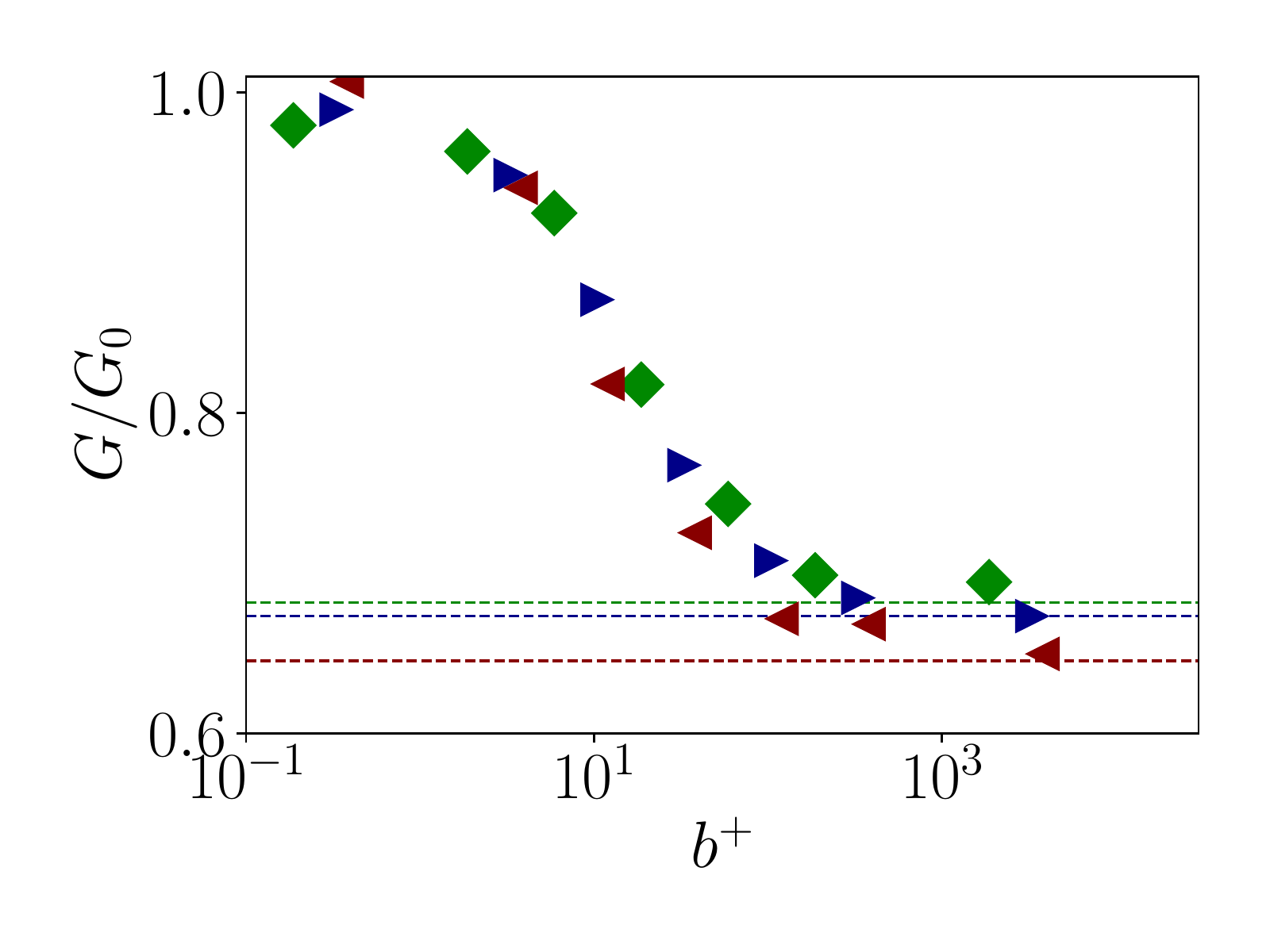}
\caption{Torque $G$ for SHP $\lambda_z=1.2$ pattern for various dimensionless slip lengths normalized using the reference no-slip case for $Re_i=10^4$ (\protect\greendia), $Re_i=2\times10^4$ (blue right-facing triangles) and $Re_i=3\times 10^4$ (red left-facing triangles). On the left, the curves are plotted against $b$ the slip length normalized using the gap-width, while on the right, they are plotted against $b^+$, i.e.~the slip-length non-dimensionalized using the viscous unit. The horizontal dashed lines represent the free-slip limit (perfect hydrophobicity) for each Reynolds numbers, while the solid curves are sigmoid fits. }
  \label{fig:sim_resk}
\end{figure}

We can support the claim that the relevant length-scale to the treatment's effectiveness is the viscous wall-unit by repeating the parameter sweep of $b$ for two additional values of $Re_i$, i.e.~$Re_i=2\times10^4$ and $Re_i=3\times10^4$. In \Cref{fig:sim_resk}, we show $G/G_0$ against $b$ and $b^+$ for all values of $Re_i$ simulated. When plotted against $b$, we can clearly see how fitting a sigmoid curve through the data results in a different inflection point, which is smaller for larger $Re_i$. However, when plotted in terms of $b^+$, i.e.~when the slip-length is non-dimensionalized using $\hat{\delta}_\nu$, the results collapse much better onto a single curve even if the asymptotic value of $G/G_0(b\to\infty)$ is different for different $Re_i$. This supports our hypothesis that the relevant parameter that controls the treatment's effectiveness is $b^+$. 

\section{Summary}\label{sec:types_paper}
To conclude, we state that the ability of superhydrophobic coating to weaken the secondary flows depends on two key factors: 

1) As seen in \cite{jeganathan2021controlling}, the best treatment pattern depends on the natural length scales of the flow. Although these can be precisely fixed in simulations through the use of small domains, they are often difficult to fix a priori in experiments because of the multiple states available to the system. This can drastically reduce the effectiveness of the treatment, as the wavelength mismatch generated by the treatment may not interfere destructively with the existing secondary flow if the length-scales do not couple. This has been shown by the difference between flat and stepped SHP surface experiments. Using the flat pattern, we have had difficulty weakening the rolls due to the different roll sizes achieved in the system. However, the stepped SHP pattern has fixed the roll aspect ratio, and we can therefore achieve an SHP pattern that drastically affects Taylor rolls. Special care must be taken to fix the flow length scale, as even changes of 20-30\% in the characteristic length scales of the secondary flow are enough to reduce, or even reverse, the effectiveness of the treatment, as shown in \Cref{fig:exp_trsigma}.

2) Another important factor that determines the effectiveness of SHP treatment in turbulent flows is not so much the physical size of the slip length but its dimensionless magnitude in viscous wall units. We have shown that, for a coating to be effective, the dimensionless slip length of the coating should be greater than a viscous wall unit ($b^+>1$). This can happen by either having a sufficiently large slip length of the SHP coating or by making the viscous length significantly smaller. This last point leads to the counterintuitive result that SHP treatments should work better at \emph{higher} Reynolds numbers, as long as they can retain stable air pockets that cause hydrophobicity within its asperities and mechanically survive the imposed shear rates. This last point merits further investigation, as it can open the door to achieving drag reduction and flow control in many types of turbulent wall flows such as pipes or boundary layers by the application of commercially-available treatments. 

In the more modest context of Taylor-Couette flows, there are other possible and more direct extensions of this work. Firstly, the coatings on the cylinders could be made to have spatially-varying strengths, and this could have a possible effect on the fixing of the roll sizes. This can be done with our current numerical code, but experimentally it presents challenges. One possibility would be to use different hydrophobic coatings. Finally, another line of research would be using hydrophobic coatings in experimental studies in the linearly-stable regime, where the secondary flows arising from the end plates initiate the transition to turbulence. By using these treatments, the effect of end plates could be mitigated. 

\textbf{Acknowledgments:} We thank the Research Computing Data Core (RCDC) at the University of Houston for providing computing resources. We acknowledge funding from the National Science Foundation through grant NSF-CBET-1934121.

\textbf{Declaration of Interests}. The authors are in the process of filling out a patent for parts of this work.

\definecolor{blue}{RGB}{34,34,255}
\newcommand{\bluecirc}{\raisebox{0.5pt}{\tikz{\node[draw,scale=0.5,blue,circle,fill=blue](){};}}}

\definecolor{yellow}{RGB}{204,204,0}
\newcommand{\yellowsq}{\raisebox{0pt}{\tikz{\node[draw,scale=0.5,yellow,regular polygon, regular polygon sides=4,fill=yellow](){};}}}

\appendix
\section{Slip Length Measurement of Superhydrophobic Surfaces using Rhemoeter}
\label{sec:rheo_sup}
To model the solid-liquid interface in macroscopic flows, it is generally assumed that the slip length $\hat b=0$, as it is in the $\mathcal{O}$ (nm) range, corresponding to the mean free path of the fluid \citep{maxwell1879vii,landau2013fluid}. However, the slip length cannot be neglected while modeling flow over stress-reducing surfaces such as superhydrophobic (SHP) ones whose slip length of $\mathcal{O}$ (mm) is much larger than the mean free path of the fluid. Therefore, it becomes imperative to measure the slip length of the SHP surface flows. To do this, we turn to \cite{srinivasan2013drag}, that uses a rheometer to measure the slip length.

\subsection{Materials}
We use a thin high precision aluminum square plate of length and width of 8 cm, and a thickness of 0.5 cm as a substrate for the SHP coating. The substrate is sandblasted to obtain a relatively rough surface using \#3 glass beads ($\approx0.85$ mm diameter). This ensures an adequate grip required to bond the SHP coatings. The SHP treatment (Ultra-Ever Dry, UltraTech International) is applied in two spraying steps. The first step is spraying the chemical called  the `bottom coat', followed by a 30 min of curing, and finally spraying the `top coat'. The bottom coat is not superhydrophobic, but it enables bonding and self-assembly of microstructures needed for superhydrophobicity found in the top coat. After 12 hours of curing time, the SHP-treated aluminum plate is transferred to a rheometer for slip length measurement.

\begin{figure}
\centering
\includegraphics[width= 0.9\textwidth]{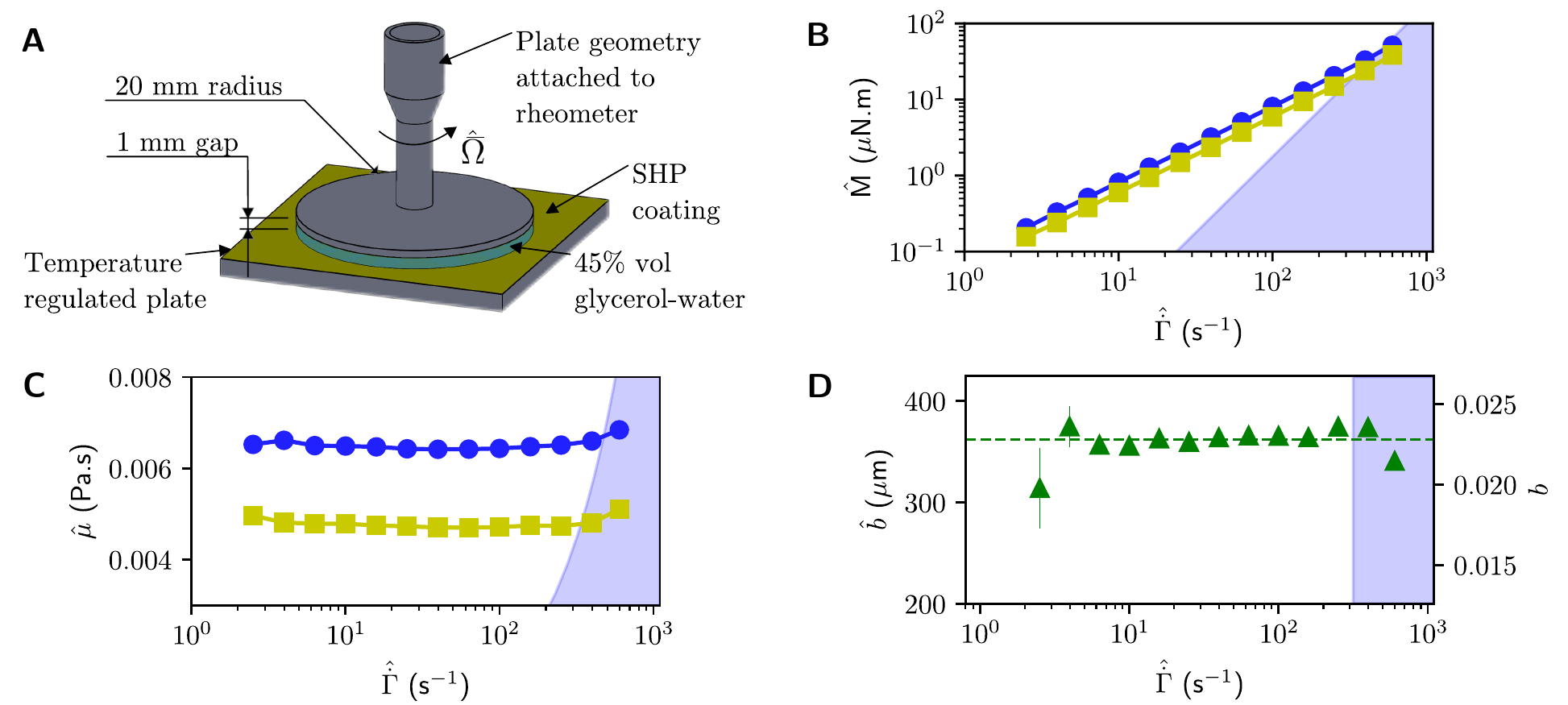}
\caption{ (\textbf{A}) Schematic of the flat plate geometry mounted on a rotational rheometer used to quantify slip length using a 45 \% glycerol–water mixture by volume as the probe liquid. (\textbf{B}) Torque measured by the rheometer over the untreated \protect\bluecirc\ and the SHP surface \protect\yellowsq\ at different shear rates. (\textbf{C}) Actual and apparent viscosity of the mixture at different shear rates measured for the untreated and SHP surfaces, respectively. (\textbf{D}) Slip length \protect\greentri\ of the SHP surface measured in $\mu$m and non-dimensionalized by the cylinders gap $\hat d$ of TCF experiments, at different shear rates (right axis). The dashed green line is the average slip length ($\hat{b}$=360$\mu$m) calculated across the shear rates (left axis). The error bars show the standard error of mean of slip lengths calculated by repeating the experiment three times. The blue regions show the unreliable zone corresponding to formation of secondary turbulent flows \citep{ewoldt2015experimental,mitra2020characterization}. }
\label{fig:rheo}
\end{figure}

\subsection{Methods}
Fig. \ref{fig:rheo} A shows the schematic of the experimental set-up. The SHP-treated aluminum plate is placed on a temperature-controlled Peltier plate maintained at 21 \textcelsius. A plate geometry of radius, $\hat R$= 20 mm, is attached to a rheometer- HR-3 Discovery Hybrid model, TA Instruments, and placed at a distance of $\hat h$= 1 mm from the aluminum plate. The gap between the SHP surface and the plate geometry is filled with 45 \% glycerol-water mixture by volume. Various rotational speeds, $\hat{\Omega}$, corresponding to a tip shear rate $\hat{\dot{\Gamma}}=\hat{R}\hat{\Omega}/\hat{h}$ in the range of $2$ s$^{-1}<\hat{\dot{\Gamma}}<600$ s$^{-1}$, are applied to the enclosed fluid. The resulting torques, $\hat {\textmd M}_{slip}$, and the derived viscosities, $\hat \mu_{slip}$, are recorded for 3 min. These are related through the formula \citep{chhabra2011non}:

% Chhabra, Raj P., and John Francis Richardson. Non-Newtonian flow and 	applied rheology: engineering applications. Butterworth-Heinemann, 2011.

\begin{equation}
 \hat{\mu} = \frac{2\hat{M}\hat{h}}{\pi\hat{\Omega}\hat{R}^4} = \frac{2\hat{M}}{\pi\hat{\dot{\Gamma}}\hat{R}^3} 
\end{equation}

Note that the viscosity measured is not the actual viscosity of the fluid but the apparent viscosity caused by the surface modification. To find the actual viscosity, $\hat \mu_{no\ slip}$, of the fluid and its corresponding torque, $\hat{\textmd M}_{no\ slip}$, the experiment is repeated with a plate that does not have the SHP coating for the same range of shear rate. The slip length, $\hat b$, of the SHP coating is then calculated for each tip shear rate using Eqn. \ref{eqn:slipl} \citep{srinivasan2013drag} :
 
\begin{equation}
\frac{\hat {\textmd M}_{no\ slip}}{\hat {\textmd M}_{slip}}=\frac{\hat \mu_{no\ slip}}{\hat \mu_{slip}}=1+\frac{\hat b}{\hat h}. \label{eqn:slipl}
\end{equation}

\noindent The experiment is repeated three times to verify repeatability and to report the standard deviation of the measurements.

\subsection{Results}
Fig. \ref{fig:rheo} B shows that the torques obtained using a rheometer for the SHP-coated plate are lower than that of the uncoated plate. This is expected since the SHP coating repels water, leading to a smaller shear stress compared to that of the uncoated surface. This lower torque also means that the viscosity calculated by the rheometer for the SHP surface is not the actual viscosity of the fluid but an \emph{apparent} viscosity. This is clearly shown in Fig. \ref{fig:rheo} C, where the viscosity calculated from the SHP plate is reduced when compared with the actual viscosity of the working fluid captured by the uncoated plate experiment. Fig. \ref{fig:rheo} D shows the slip length calculated using Eqn.~\ref{eqn:slipl}, averaged across three repetitions of the same experiment. The average slip length across all the shear rates is found to be $\hat b=360 \pm 12\ \mu $m. This corresponds to $b=\hat{b}/\hat{d} \approx 0.023 $ in dimensionless terms, where $\hat d$ is the gap width used in the Taylor-Couette flow (TCF) experiments. The blue regions in Figs.~\ref{fig:rheo} B-D show the onset of turbulence corresponding to the critical Reynolds number $Re_{crit} \approx 4$ \citep{ewoldt2015experimental,mitra2020characterization}, where the data are not reliable. The maximum reliable viscosity based on this critical Reynolds number and its corresponding torque that can be measured by rheometer are given by Eqns.~\ref{eqn:rel_vis} and \ref{eqn:rel_tor} respectively:

\begin{equation}
\hat \mu> \frac{\hat \rho \hat{h}^3}{\hat R Re_{crit}} \hat {\dot{\Gamma}}, \label{eqn:rel_vis}
\end{equation}

\begin{equation}
\hat {\textmd M}> \frac{\pi  \hat \rho \hat R^2 \hat h^3}{2 Re_{crit}} \hat {\dot{\Gamma}}^2. \label{eqn:rel_tor}
\end{equation}

\noindent $\hat \rho$ in the Eqns.~\ref{eqn:rel_vis} and \ref{eqn:rel_tor} represents the density of the working fluid. 

\section{Shear Test of SHP Samples} \label{sec:sh_test}
SHP treatments cause superhydrophobicity on a surface by trapping air in the asperities of myriad of micro and nanostructures. The durability of SHP coating hinges on the strength of the bonds that attach these structures to the substrate. When adequate stress is applied, these bonds could be overcome, leading to the removal of microstructures and the loss of superhydrophobicity. This is true especially under extreme conditions such as high pressure-driven channel flows that cause enormous stress on the walls, where the SHP surfaces are largely implemented. Hence, there are a variety of studies \citep{wang2016transparent,xue2015fabrication} that conduct abrasion studies to assess the durability of coatings. However, these tests are not scenario-specific and any conclusion derived from these studies regarding the durability of the SHP coating in our TCF experiments would require substantial approximation. Therefore, we conduct specific shear tests on our SHP coating to study its durability.

\subsection{Materials and Methods}
Eight samples of 1 cm length by 1 cm width are cut using a 0.25 cm thick aluminum plate. The samples are sandblasted and SHP treated using the technique described in the previous section. The samples are fixed on the inner cylinder of the TCF experimental set-up using a double-sided adhesive tape. We fill the gap between the inner and outer cylinders with demineralized water, the same working fluid used to perform all TCF experiments, and shear the samples for $Re_i=2\times 10^4$ corresponding to a shear rate of $\hat{\dot{\Gamma}}=600$ s$^{-1}$. The samples are subsequently removed at certain intervals, to creating a range of specimens, each sheared only for a certain duration. ImageJ \citep{schneider2012nih} is used to measure the contact angles of 5 $\micro$L deminearlized water droplet snapshots captured with a Phantom VEO 710 camera.

\subsection{Results}
\begin{figure}
\centering
\includegraphics[width=0.9\textwidth]{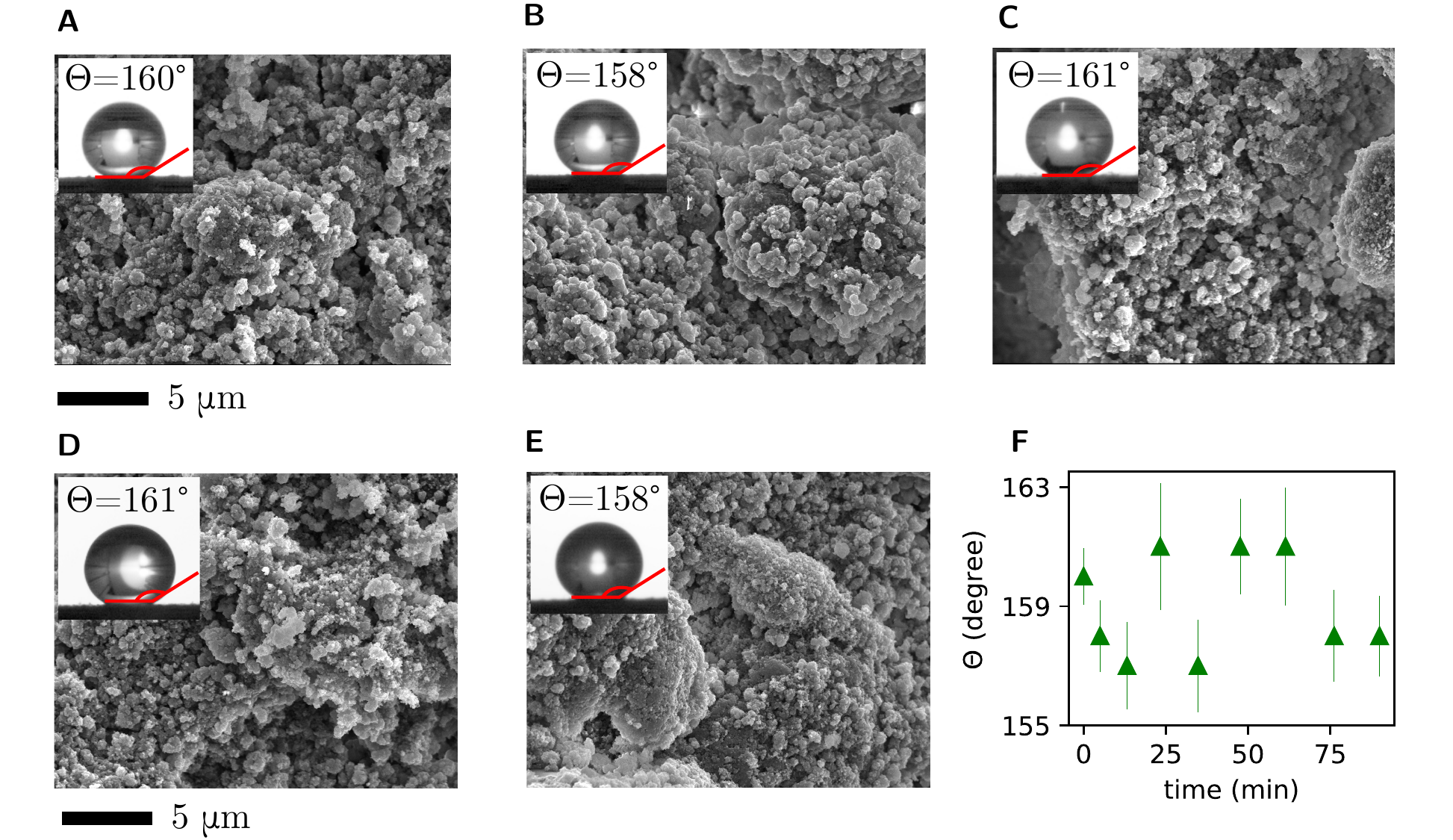}
\caption{ SEM images of the SHP surface under various durations of shear rate $\hat{\dot{\Gamma}}=600 $ s $^{-1}$ : \textbf{A}) 0 min (\textbf{B}) 5 min, (\textbf{C}) 20 min, (\textbf{D}) 60 min, and (\textbf{E}) 90 min. The insets show the contact angle of a $5 \ \mu$L demineralized water droplet over the corresponding surfaces. (\textbf{F}) The contact angle \protect\greentri\ of SHP surfaces at different durations of shear rate $\hat{\dot{\Gamma}}=600 $ s $^{-1}$. The error bars show the standard deviation of the contact angle obtained during the elliptical curve fitting of the droplets.}
\label{fig:stress}
\end{figure}

Figs.~\ref{fig:stress} A-E show the SEM images of the freshly coated SHP sample and those corresponding to various durations of shear rate $\hat{\dot{\Gamma}}=600$ s$^{-1}$. It is clear that the SHP coating is durable at the shear rate studied, since all of the samples observed retain microstructures that cause superhydrophobicity. Further proof of superhydrophobicity is seen in the insets of these images, which shows that the contact angle has remained as high as $\Theta = 159\degree \pm 2 \degree$. The contact angles are plotted against different shear durations of shear rate in Fig.~\ref{fig:stress} F.

\FloatBarrier

\section{Torque and Velocity Benchmarks} 
\label{sec:expcomp}
\begin{figure}
\centering
\includegraphics[width=0.45\textwidth]{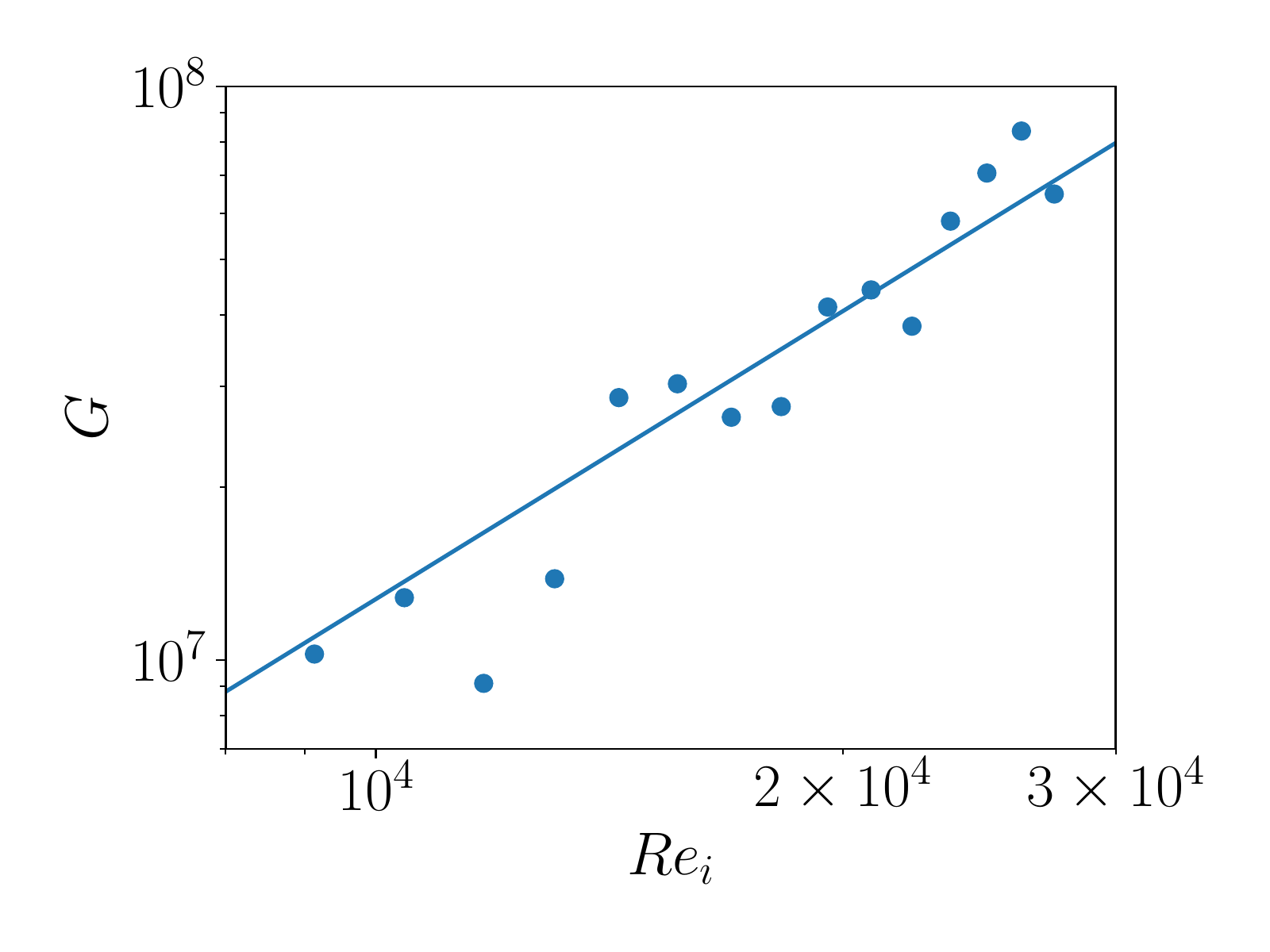}
\includegraphics[width=0.45\textwidth]{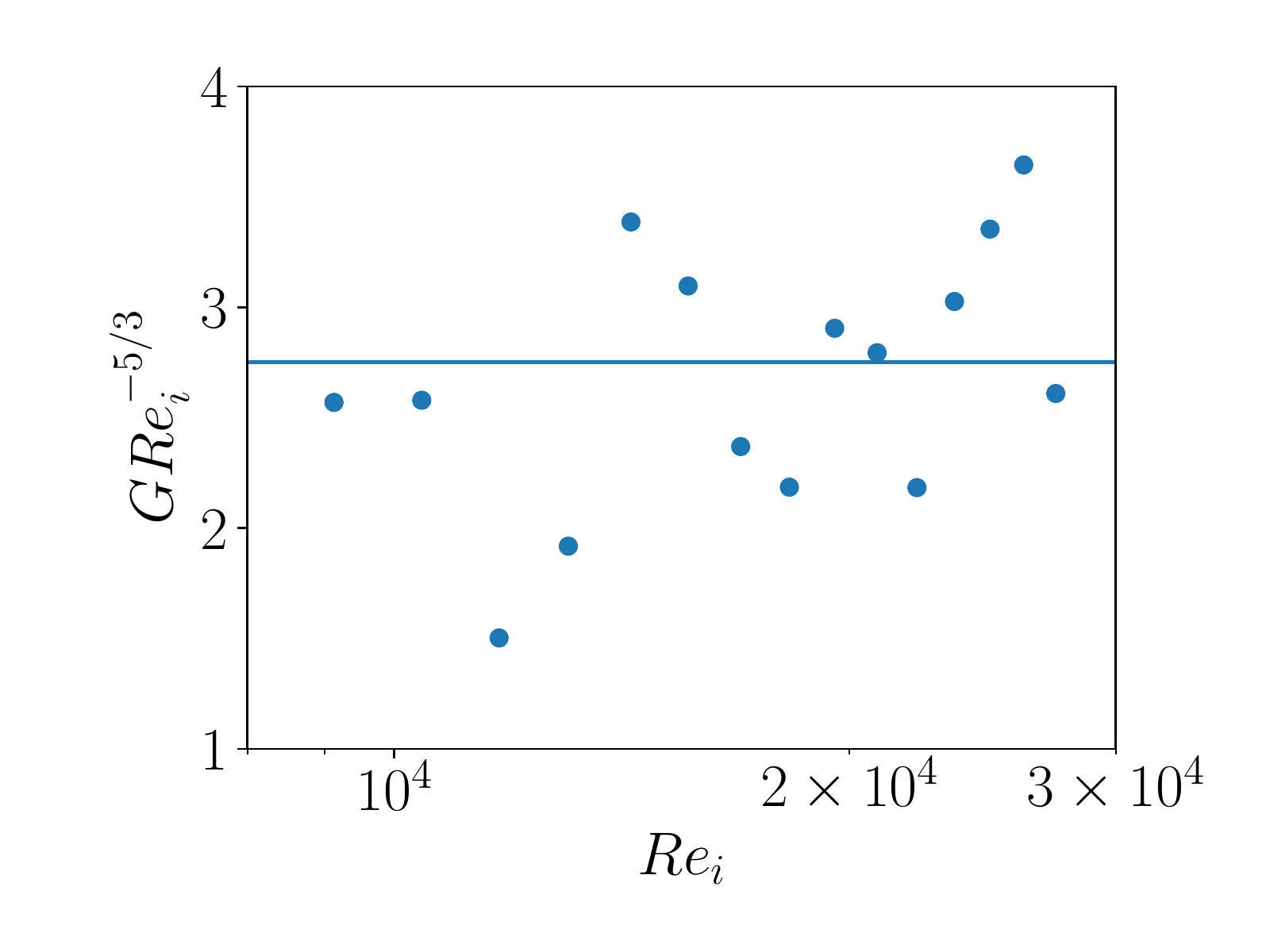}
\caption{ Uncompensated and compensated non-dimensional torque $G$ measurements against $Re_i$ for the untreated cylinders. The solid line is $G=0.202(\eta^{-1}-1)^{-5/3}Re_i^{5/3}$ \citep{marcus1984simulation,lathrop1992transition}.  }
\label{fig:expcomp}
\end{figure}

In this section, we present a series of benchmarks for our experimental set-up. In \Cref{fig:expcomp}, we show uncompensated and compensated torques against $Re_i$. The data for these measurements is obtained by subtracting the torque measured when the cylinders are filled with water from the torque measured when the cylinders are filled with air. The latter number provides a reference estimate for the losses in the system. We add the benchmark $G=0.202(\eta^{-1}-1)^{-5/3}Re_i^{5/3}$ from \cite{marcus1984simulation,lathrop1992transition}. The torque measurements show a degree of dispersion around the benchmark line even though they largely follow the trend.

\begin{figure}
\centering
\includegraphics[width=0.45\textwidth]{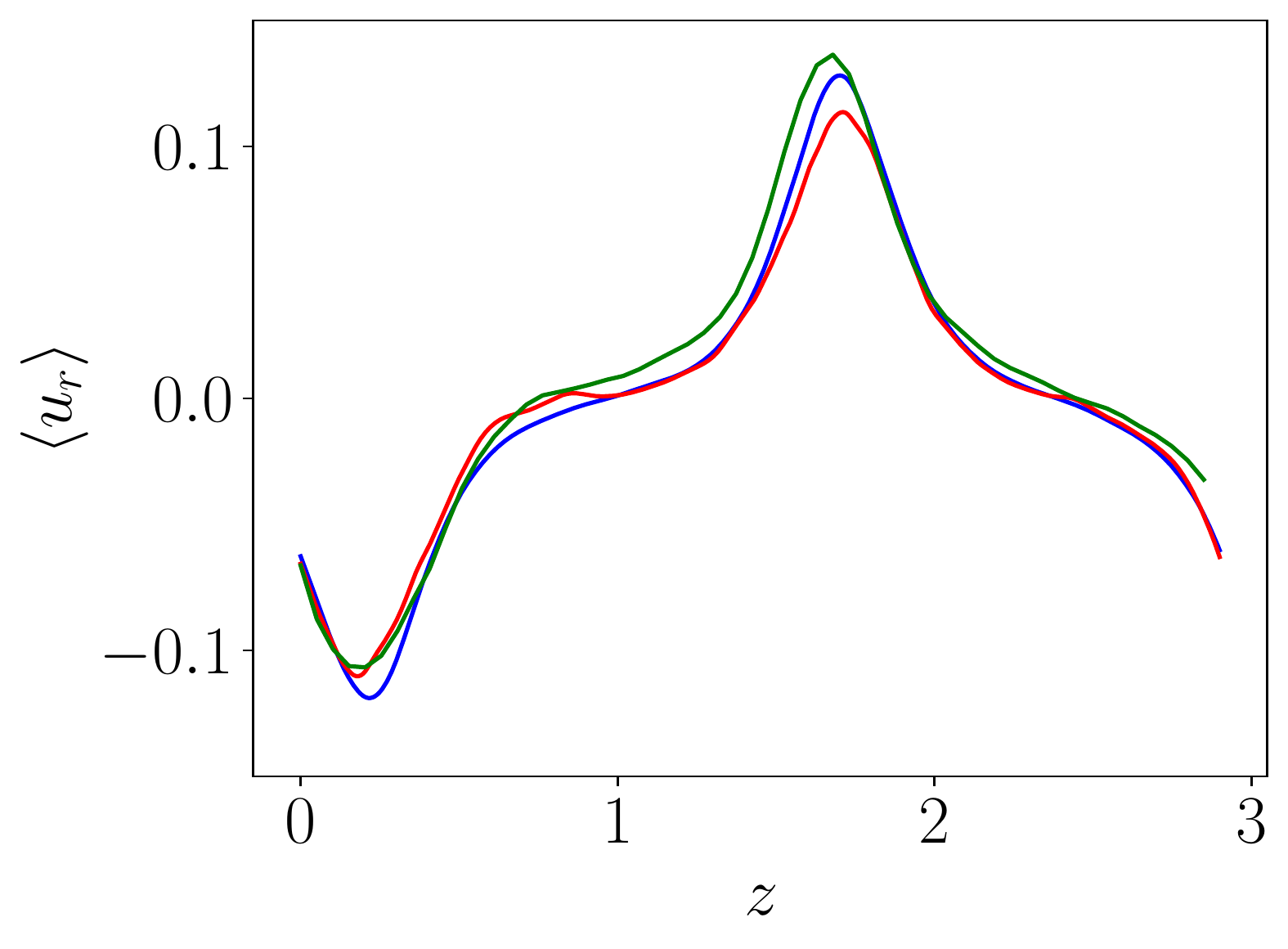}
\includegraphics[width=0.45\textwidth]{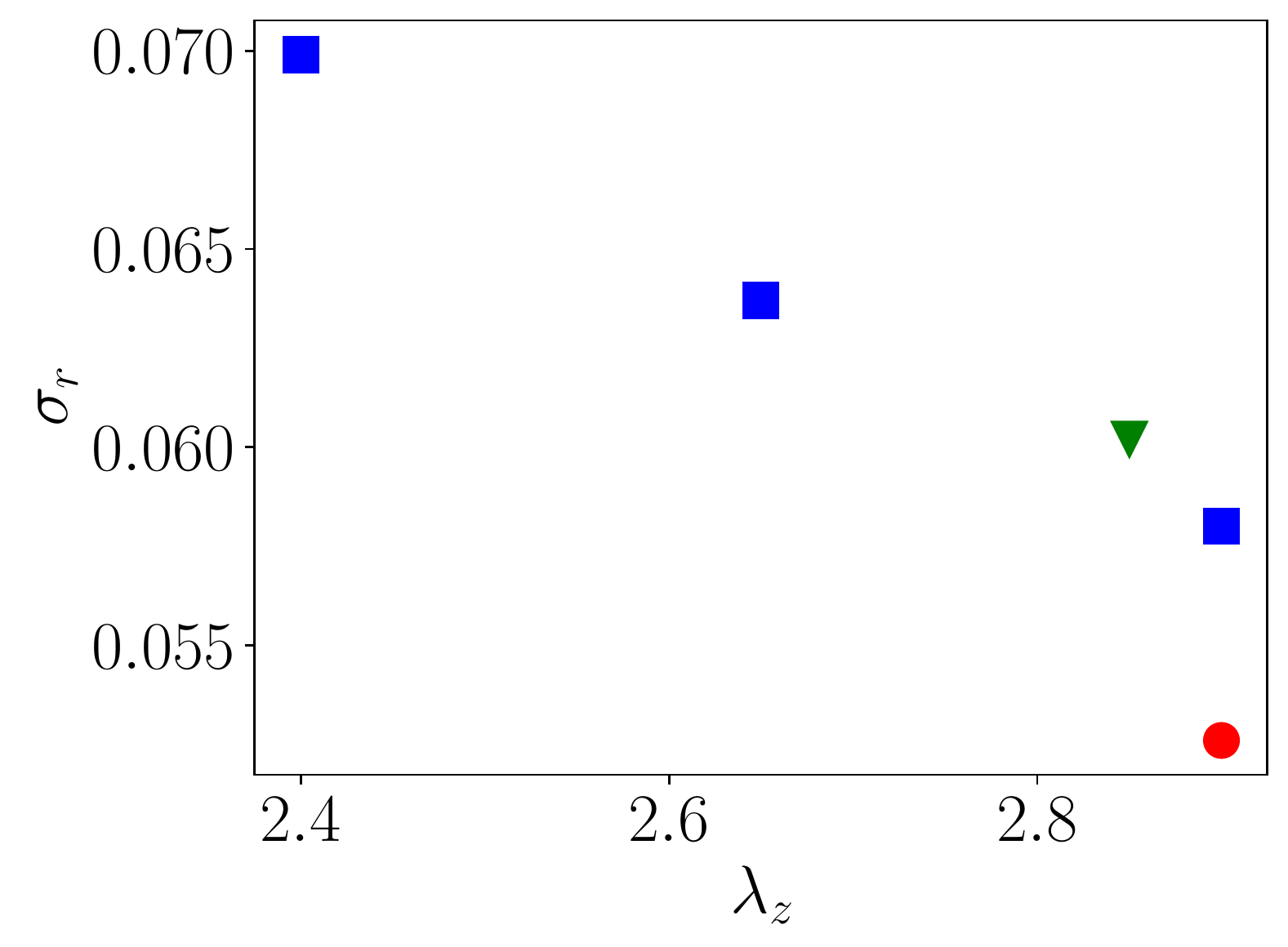}
\caption{ Left: comparison of the radial velocity at the mid-gap obtained from experiments (green) and simulations (blue for $\langle u_r \rangle_{t,\theta}$, red for $\langle u_r \rangle_t$) for untreated cylinders and $\lambda_z\approx 2.9$. Right: values of roll strength $\sigma_r$ obtained from $\langle u_r \rangle_t$ for simulations (red circle) and experiments (green triangle), and from $\langle u_r \rangle_{t,\theta}$ for simulations (blue squares) at $Re_i=10^4$ for untreated cylinders. }
\label{fig:app_velocity}
\end{figure}

In \Cref{fig:app_velocity}, we compare the velocity data obtained from the PIV performed in experiments to that obtained from simulations. In \Cref{fig:app_velocity}A, we show the experimental results for $\langle u_r\rangle_t$ at the mid-gap in green, as well as the values for $\langle u_r \rangle_{\theta,t}$ obtained from simulations at a similar $\lambda_z$ in blue. We also show simulation data obtained by only averaging temporally and not azimuthally in red, i.e.~$\langle u_r \rangle_t$. We can observe that both procedures for obtaining the average radial velocity in simulations give results which have a similar axial profile as the experiments. The data obtained from simulations by only averaging azimuthally, i.e.~$\langle u_r \rangle_t$, appear less statistically converged due to the smaller amount of samples. In \Cref{fig:app_velocity}B, we show the value of $\sigma_r$ calculated from the no-slip experiment, as well as the value obtained for $\sigma_r$ from several simulations by using $\langle u_r \rangle_{\theta,t}$ and $\langle u_r \rangle_t$. The numerical procedure for calculating $\sigma_r$ from $\langle u_r \rangle_{\theta,t}$ results in values of $\sigma_r$ which match quite well the data obtained from experiments. The value of $\sigma_r$ obtained from $\langle u_r \rangle_t$ in simulations is slightly lower than the other values. The data also show the trend of $\sigma_r$ with $\lambda_z$ described in \Cref{sec:expresults}.

\FloatBarrier

\section{Supplementary Movies} \label{sec:vid}

\textbf{Movie M1.} Particle Image Velocimetry (PIV) experiment of no-slip Taylor-Couette flow (TCF). The left and right edges are the inner and outer cylinders respectively. The video is slowed 80 times.

\textbf{Movie M2.} PIV experiment of TCF with stepped superhydrophobic (SHP) coating of pattern wavelength $\lambda_z=1.2$ on the inner cylinder. The left and right edges are the inner and outer cylinders respectively. The video is slowed 80 times.

\textbf{Movie M3.} Direct Numerical Simulation (DNS) video of instantaneous non-dimensional radial velocity of the no-slip TCF. The left and right edges are the inner and outer cylinders respectively. The video is slowed 80 times.

\textbf{Movie M4.} DNS video of instantaneous non-dimensional radial velocity of the TCF with free-slip pattern of wavelength, $\lambda_z=1.2$, on the inner cylinder. The left and right edges are the inner and outer cylinders respectively. The video is slowed 80 times.

\textbf{Movie M5.} DNS video of instantaneous non-dimensional radial velocity of the TCF with finite-slip pattern of wavelength, $\lambda_z=1.2$, and slip length, $b=0.023$, on the inner cylinder. The left and right edges are the inner and outer cylinders respectively. The video is slowed 80 times.

\bibliographystyle{jfm}
\bibliography{jfm}

%% End of file `jfm2esam.bib'.

\end{document}